\documentclass[a4paper,11pt]{article}
\usepackage{jcappub} % for details on the use of the package, please see the JINST-author-manual
\usepackage{lineno}
\usepackage{amsmath}
\usepackage{amsfonts}
\usepackage{graphicx}
\usepackage{cancel}
\usepackage{stmaryrd}
\usepackage{graphicx}
\usepackage{float}
\usepackage{caption}
\usepackage{subcaption}
\usepackage{pifont}
\usepackage{booktabs}
\usepackage{multirow}
\usepackage{enumitem}
\usepackage{xcolor} % needed for \yl/\tc/\pg colored margin notes (was previously pulled in transitively by nicematrix)

\newcommand{\aK}{\alpha_K}
\newcommand{\aB}{\alpha_B}
\newcommand{\aH}{\alpha_H}
\newcommand{\aT}{\alpha_T}
\newcommand{\aM}{\alpha_M}
\newcommand{\abar}{\bar{\alpha}}

\newcommand{\Mpl}{{m}_*}
\newcommand{\R}{{\cal R}}
\newcommand{\cE}{{\cal E}}
\newcommand{\cN}{{\cal N}}
\newcommand{\Src}{{\cal S}}

\renewcommand{\d}{{\rm{d}}}
\newcommand{\yl}[1]{#1}
\newcommand{\ylb}[1]{#1}

\title{Stochastic scalar-tensor inflation and beyond}
\author[a]{Yoann L. Launay,}

 \affiliation[a]{Centre for Theoretical Cosmology, Department of Applied Mathematics and Theoretical Physics,
University of Cambridge, Wilberforce Road, Cambridge CB3 0WA, United Kingdom}
\emailAdd{yoann.launay@outlook.com}

\abstract{
During cosmological inflation, inhomogeneities arising from quantum vacuum fluctuations are stretched to become super-Hubble and effectively classical. As many scenarios of the origin involve nonlinearities or a breakdown of perturbativity in the infrared, the limitations of quantum field theory can be addressed using a stochastic description of the dynamics, the so-called stochastic inflation paradigm. However, the stochastic formalism was only recently formulated consistently within full General Relativity and has not yet been extended to more general theories of the early universe, which is the subject of this work.

In order to find the stochastic sources for a wide class of fully nonlinear scalar-tensor theories, we apply our gauge-agnostic coarse-graining procedure to the linear equations of the \yl{associated effective field theory}. Each theory can then be mapped to its own set of stochastic equations of motion by identifying the corresponding coefficients in the \yl{former}. We illustrate this with a few concrete and, in most cases, unprecedented examples, including Gauss-Bonnet, generalized Brans-Dicke, Horndeski, and braiding theories.

Finally, we discuss other natural extensions to provide a phenomenologically complete stochastic framework. For example, we showcase the coarse-graining of multifield inflation in full General Relativity and argue for the generality of our procedure and thus its potential applications beyond the realm of inflation.
}

\begin{document}
\maketitle
\flushbottom

\section{Introduction}
% \pg{A diagram for the procedure would help. EFT, stochastic, modified gravity, etc.}
Inflation is the most widely accepted formalism for the description of the early universe \cite{starobinsky80,guth_1981,Linde82}, providing a compelling explanation for the observed homogeneity, isotropy, and flatness of the cosmos, as well as a mechanism for generating the primordial density perturbations that seeded the formation of large-scale structure \cite{Mukhanov1981,Hawking1982,Starobinsky1982pert,Guth82pert,HawkingMoss1982,Bardeen83}.
However, the dynamics of inflation in this framework are believed to be inherently quantum mechanical from the vacuum fluctuations of the inflaton field, and are still an active area of research, with, as mentioned below, many remaining open questions and challenges.

Stochastic Inflation, born not so long after Inflation \cite{Starobinsky_stochastic_1988}, has proved itself as the main complementary method to the standard quantum field theoretic approaches in the prediction of observables.
As it is possible to observe the breaking of perturbativity in some scenarios, stochastic inflation makes the most of the seemingly coincidental classicality there to provide a nonperturbative model of the dynamics, semiclassically sourced by the quantum vacuum fluctuations\textquoteright{} noise. 
Since its early days \cite{Starobinsky_stochastic_1988,salopek_nonlinear_1990,salopek_stochastic_1991,rigopoulos_non-linear_2005,vennin_correlation_2015}, Stochastic Inflation has been used in the long wavelength limit of General Relativity (GR), yielding numerous developments and predictions on classicalised primordial perturbations \cite{nakao_stochastic_1988,
nambu_stochastic_1988,
sasaki_classical_1988,
hosoya_stochastic_1989,
habib_stochastic_1992,
bellini_stochastic_1996,
Winitzki00,
rigopoulos_separate_2003,
liguori_stochastic_2004,
tsamis_stochastic_2005,
Rigopoulos_multi06,
BattefeldVanTent06,
tolley_stochastic_2008,
Finelli09,
Finelli10,
gratton_path_2011,
Fujita2013,
levasseur_lagrangian_2013,
levasseur_recursive_2013,
rigopoulos_fluctuation-dissipation_2013,
garbrecht_infrared_2014,
burgess_eft_2015,
levasseur_backreaction_2015,
assadullahi_multiple_2016,
burgess_open_2016,
grain_stochastic_2017,
pattison_quantum_2017,
vennin_critical_2017,
prokopec_functional_2018,
pattison_attractive_2018,
noorbala_tunneling_2018,
markkanen_scalar_2019,
pinol_inflationary_2019,
firouzjahi_stochastic_2019,
pattison_stochastic_2019,
ezquiaga_exponential_2020,
Bounakis20,
de_numerically_2020,
pattison_ultra-slow-roll_2021,
ando_power_2021,
cohen_stochastic_2021,
cable_free_2021,
prokopec_ensuremathdeltan_2021,
Jackson22,
tasinato_stochastic_2022,
Figueroa22,
tada_statistics_2022,
cable_second-order_2022,
cruces_stochastic_2022,
mishra_primordial_2023,
wilkins_stochastic_2023,
tomberg_numerical_2023,
Tomberg:2024evi,
tajima_stochastic_2024,
cruces_small_2024,
jackson_stochastic_2024,
ballesteros_non-gaussian_2024,
noorbala_classicality_2024,
sharma_stochastic_2024,
Animali25,
tasinato_stochastic_2025,
nassiri-rad_stochastic_2025,
miyamoto_calculating_2025,
briaud_stochastic_2025,
blachier_friction_2025,
barenboim_large_2025,
murata_stochastic_2026,
saha_nonlinear_2026,
beneke_quantum_2026,
ye_nonperturbative_2026,
cruces_consistent_2026,
mishra_eigenvalue_2026,
animali_time-reversed_2026,
Li26b}.
Some of these refinements of Starobinsky's original idea over the years have arised to describe more and more regimes.
In particular, many works explored the breaking of perturbativity beyond secular growth arising on fixed dS backgrounds.
In general with a full spacetime, \yl{whether or not a theory can produce such secular effects \cite{Assassi12}, this does not protect it from nonperturbative dynamics.}
For instance, models of ultra-slow-roll phases of inflation have attracted significant interests in their promise to produce strong overdensities and subsequent primordial black holes.

Among others, such models and methods have also shown some challenges given the subtleties and debates in the full account of gradients, rightful gauges, backreaction etc, see for instance \cite{artigas_hamiltonian_2022, Artigas25, jackson23}.
Recent work \cite{Launay24} shows that, if one is ready to give up analytical solutions, it is possible to build such a formalism in full general relativity, and to solve it numerically \cite{Launay26}, without having to think about the validity of many historical approximations apart from classicality (see \cite{Launay2024bis} on the matter).
The procedure employed there is deeply rooted in the mastery of linear perturbation theory in any gauge, and, as conveyed in this work, can be extended beyond single scalar-field inflation in GR.

With the same minimal approximations, this work provides the generalisation of stochastic inflation to scalar-tensor theories going beyond General Relativity.
These theories constitute in fact the generic extension of GR with a scalar degree of freedom relevant for inflation and, more generally, dark energy, and encompass a wide class of viable inflationary frameworks while remaining theoretically controlled.
In the perturbative limit ---useful to derive stochastic sources as we will recall--- these theories are easierly gathered in the same formalism using the Effective Field Theory of Inflation (EFToI)\ylb{.
Initially sparkling from the EFToI, the EFT of Dark Energy (EFToDE) extended the framework to generic dark energy in the past two decades, as part of an effort to constrain post-inflation models}, and even saw some of the parameters ruled out for late eras of cosmology \cite{CreminelliVernizzi17}.
As we will see, the associated notations, in particular those of the so-called $\alpha$ basis \cite{BelliniSawicki14}, are still particularly meaningful for the study of inflation.
The EFToI is today a subset of this scalar-tensor framework, and usually goes further in its approximations.
In particular, mixings with gravity are usually neglected by assuming that the Hubble parameter, or the energy, is above the decoupling scale (equivalent in practice to $\Mpl\to \infty$, and $\dot H \to 0$, at fixed $\Mpl^2\dot H$).
In the first place, these energies are much higher than the late-time ones but the freezing of the gauge-invariant fluctuations at Hubble scale crossing makes the imprint at those energies relevant.
The predictability of such approach has recently culminated thanks to perturbative solvers such as $\mathsf{CosmoFlow}$ \cite{werth23}.  

For the case of inflation and before even mentioning nonperturbative effects, the perturbative approach is not without generalisations and motivations beyond the decoupling limit.
In particular, computations of the curvature perturbation 3-point function ---as the leading signal of inflationary interactions--- have been performed beyond this limit.
The pioneering work of Maldacena \cite{Maldacena_2003} gave early on the 3-point function of the curvature perturbation in single-field slow-roll inflation with the mixings with metric perturbations and the leading slow-roll contributions in General Relativity.
This was later extended to more general single-field actions\footnote{As well as various multifield extensions, including the decoupled EFToI treatment of \cite{senatore_effective_2012}.}, such as those with single-field non-canonical kinetic terms \cite{seery_primordial_2005, chen_primordial_2010, burrage_large_2011}, including braiding \cite{zhang_primordial_2021}, but also various calculations in the general Horndeski classes with those mixings involved \cite{de_felice_inflationary_2011, felice_primordial_2011}, including the gravitonic bispectra \cite{gao_primordial_2011}.
To make such tedious calculations accessible in non-decoupled regimes, numerical pipelines such as \cite{Zhang25} have also been developed.
Importantly, these works emphasise that nontrivial contributions can be found and be missed by the decoupling limit in a generic scalar-tensor theory.
Moreover, even if GR or the decoupling limit are found to be good approximations to certain energy scales, such extended treatments are necessary to quantify the validity of the former.
It is also worth mentioning that a plethora of inflationary models formulated in GR, whether given as interactive actions or simple potentials (e.g. the widespread \textit{scalaron} Starobinsky model \cite{starobinsky80}), have been derived from scalar-tensor theories using transformations to the so-called \textit{Einstein frame}.
However such transformations between frames do not always exist \cite{Zumalacarregui13} or could generate increased complexity beyond the single-field scenario, as a field space curvature might be generated in the multifield case \cite{Kaiser10}.

In the aforementioned existing calculations, beyond the decoupling limit or not, several approximations are employed on top of perturbativity and the tree-level limitation\footnote{Loop calculations are still challenging today \cite{Senatore10,Assassi12,lee_leading_23} and are thus not part of any numerical pipeline.}, such as the freezing of couplings at the horizon or perturbative expansions on state properties or couplings (including sound speed, slow-roll parameters etc).
More generally, computations of correlators ---even at tree-level--- have never been implemented for a generic scalar-tensor theory, whether using in-in integrations or abovementioned transport equations.
In this context, one can see that the fully quantum methods are not only unable to handle nonperturbative dynamics at the moment, but also to handle systematically the full variety of perturbative and nonlinear dynamics in the landscape of inflationary models.

At the expense of analytic work and most importantly quantum effects, relying on evolving the equations of motion is usually the easiest way to resolve full nonlinearities and nonperturbative effects, as shown by multiple works in the context of inflation \cite{Cosmolattice,latticeeasy,clough_robustness_2017,Elley2024,Caravano22thesis,Launay2025,Giannadakis25}.
However, the stochastic approach is precisely restricted to a semiclassical regime, and so, it is not a problem to give up the offshell part of the system's description, as long as the classicality condition is satisfied.
% This what we was done in recent work in GR \cite{Launay24,Launay26}, and what we generalise here to scalar-tensor theories.
Providing a theory of stochastic inflation in a generic scalar-tensor theory is thus about providing a simpler, if not first, access to correlators from generic theories of the origin by exploiting the semiclassicality of the inflationary spacetime.

This paper is organised as follows.
The general philosophy and procedure to build stochastic equations from an IR theory is briefly introduced in Section~\ref{sec:ansatz}.
In order to derive the stochastic sources of generic scalar-tensor theories via their linear equations without applying the coarse-graining to each theory separately, we introduce the \yl{EFT action} and our chosen notations in Section~\ref{sec:EFToDE}.
The stochastic ansatz is then applied to get our main result summarised in Section~\ref{sec:mainresult}: the stochastic sources of the scalar-tensor equations of motion of theories admitting a decomposition in the \yl{EFT action} when linearised.
In Section~\ref{sec:EOMs}, we put this result into practice for a variety of theories.
This means finding the \yl{EFT action} coefficients that each chosen theory corresponds to, including nonminimally coupled theories, Einstein-Gauss-Bonnet, Horndeski, Braiding, before applying the main result.
Finally, Section~\ref{sec:discussion} concludes by pushing the stochastic ansatz to its edge discussing further possible extensions, inspired by previous works and the generality of our \yl{EFT action} notations.
For instance, we demonstrate the generality and naturalness of our advertised procedure by coarse-graining multifield GR for the first time. 

\paragraph{\yl{Notation}.}

In the following, we will use the following tensor for comparison to the Einstein field equations,
$$
  {\cal E}_{\mu\nu} = \frac{2}{M^2}\frac{1}{\sqrt{-g}}\frac{\delta {\rm S}}{\delta g^{\mu\nu}},
$$
with $M$ a potentially time-dependent Planck mass and $S$ the action of our theory.
When ${\rm S}$ is the Einstein-Hilbert action with some content described by a stress-energy tensor $T_{\mu\nu}$, we recover ${\cal E}_{\mu\nu} = G_{\mu\nu}-\Mpl^{-2} T_{\mu\nu}$, with the Planck mass $\Mpl = (8\pi G)^{-1/2}$. ${\cal E}_{\mu\nu}$ vanishes in the least action configuration (\textit{on-shell}).
In full generality, we also define 
$${\cal E}_Z = \frac{2}{M^2}\frac{1}{\sqrt{-g}}\frac{\delta {\rm S}}{\delta Z}, $$
useful for instance to look at scalar field equations of motion. \yl{$Z$ denotes a generic dynamical field or metric variable of the theory.}

We will also use ${\cal F}$ to denote the Fourier transform, and \yl{$\boldsymbol{\xi}_{\mathbf{k}}$} to denote the Fourier transform of a white noise, together with usual notations of an FLRW spacetime with scale factor $a(t)$, Hubble rate $H = \dot{a}/a$, Hubble slow-roll parameters $H \varepsilon_{n+1} \equiv - d_t{\ln \varepsilon_n}$, $t$ being the cosmic time.
Apart from $a, H, \varepsilon_n$ which are implicitly background quantities, we will use the subscript $b$ to denote the background value of a field.

\section{Scalar-tensor stochastic sources} \label{sec:procedure}

\subsection{Stochastic ansatz\label{sec:ansatz}}
Stochastic inflation\footnote{I defer to \cite{Starobinsky_stochastic_1988,vennin_correlation_2015,cruces_review_2022,Launay24} for further technical and state-of-the-art reviews.
} was originally designed to tackle the description of IR secular growth of perturbations of a $\lambda\phi^4$ inflaton in de Sitter \cite{Starobinsky_stochastic_1988}.
As IR modes were growing beyond the perturbativity bounds of QFT, it was convenient to observe the (nonperturbative-growth-friendly) classicality of each mode $k$ as the Hubble radius becomes smaller ($k\leq aH$) during inflation.
Concretely, commutators were found to be decaying from this time-dependent cutoff and quantum operators turned into stochastic numbers, with a quantum-originated spectrum.
The stochastic ansatz \yl{thus simply assumes} that, statistics-wise, inhomogeneous inflation can be looked at using the classical open IR system of scales, described by stochastic \yl{partial differential equations}\footnote{In principle, this doesn't require the noise to be originated from a quantum theory using quantum initial conditions.}.
Beyond that simple ansatz come further nuances in diverse treatments, see for instance historical approximations used in General Relativity \cite{salopek_nonlinear_1990,salopek_stochastic_1991}.
The common point of all stochastic equations of motion formulated so far for inflation, which could thus be appended to the stochastic ansatz, is the linear approximation for the UV quantum modes.
In mathematical terms, focusing without loss of generality on a field $X$, with decomposition $X = X^<+X^>$ between its IR classical and nonlinear part $X^>$ and its UV part $X^<$, this can be written with a nonzero r.h.s.
\begin{equation}
  \begin{aligned}
     \cE_{X^>} & \simeq {\cal F}^{-1}\{S_X(k)\boldsymbol{\xi}_{\mathbf{k}}\}\,,\\
    \Src_X(k) & = \cE^{(1)}_{X}[W_kX_k] \pmod{W_k\cE^{(1)}_{X}[X_k]},
  \end{aligned}
\end{equation}
where $\cE_{X^>}$ is the original classical equation of motion applied to the IR field $X^>$, and $\cE^{(1)}_{X}[X^<]$ is the original linearised equation of motion applied to the UV field $X^<$.
The linear operator uses the average of the IR field as its background (\textit{Starobinsky approximation}), or the IR field itself depending on the treatment of the stochastic backreaction.
The separation has been made explicit using a window function $W$ selecting the IR modes, $X^> = W*X$ and $X^< = (1-W)*X$.
We wrote the approximation as an equality up to the term $W*\cE^{(1)}_{X}[X](=0)$ to emphasise that the linear operator is applied to the product of the UV window and the field, which will simplify dramatically this r.h.s..
In practice, the source is computed by solving in parallel for the spectrum of the noise using $\cE^{(1)}_{X}[X_k]=0$. In this manuscript, we refer to this whole algebraic procedure as the \textit{coarse-graining}.

\ylb{We sketch here a simplistic proof for the reader.}
Let ${\rm S}$ be an action of interest, functional of one field $X$ for simplicity.

% We assume perturbativity in the UV field next, $X^< \ll X^>$, and write the functional Taylor expansion of the action up to first order
\yl{Let us expand the total action}
\begin{equation}
  {\rm S}[X^>+X^<] = \underbrace{{\rm S}[X^>]}_{{\rm S}^{(0)}}+\underbrace{\int d^4x \frac{\delta {\rm S}}{\delta X(x)}\Bigg|_{X^>}X^<(x)}_{{\rm S}^{(1)}}+O\Big((X^<)^2\Big).
  \label{eq:actionExpansion}
\end{equation}
\ylb{$X^>$ contains here a background mode justifying this expansion in amplitude, as opposed to studies of fields in the decoupling limit.
Beyond a naive expansion in the UV field against the IR field, this can be justified in explicit derivations integrating out the UV fields in the Closed-Time-Path formalism by assuming negligible UV interactions\footnote{See also first UV loop corrections to the effective potential \cite{moss_effective_2017}.} \cite{morikawa_dissipation_1990,pinol_manifestly_2021}, or using more implicit Open EFT and Renormalization Group arguments \cite{Cespedes26}.}
\yl{This allows us to extract the equation of motion (EOM) of the IR field with linear UV corrections}
\begin{equation}
  \begin{aligned}
      \frac{\delta {\rm S}}{\delta X^>(x)}\Bigg|_{X^>} & = \frac{\delta {\rm S}^{(0)}}{\delta X^>(x)}\Bigg|_{X^>} + \frac{\delta {\rm S}^{(1)}}{\delta X^>(x)}\Bigg|_{X^>}+O\Big((X^<)^2\Big), \\
                                      &  = \frac{\delta {\rm S}}{\delta X(x)}\Bigg|_{X^>} + \int d^4y \frac{\delta^2 {\rm S}}{\delta X(\ylb{x})\delta X(y)}\Bigg|_{X^>}X^<(y)+O\Big((X^<)^2\Big),
  \end{aligned}
\end{equation}
after injecting the previous expansion. 
Using now the classicality assumption on $X^>$, we can assume that the evolution is entirely described by $$\frac{\delta {\rm S}}{\delta X^>(x)}\Bigg|_{X^>}=0,$$
so that
\begin{equation}
  \frac{\delta {\rm S}}{\delta X}\Bigg|_{X^>} \yl{\simeq} - \int d^4y \frac{\delta^2 {\rm S}}{\delta X(\ylb{x})\delta X(y)}\Bigg|_{X^>}X^<(y).
\end{equation}
This says that the classical EOM of the IR field is the same as that of the original EOM, supplemented by a right-hand side source fueled by the UV field.
The double functional derivative applied to S is the inverse propagator, which is the linear equation of motion operator\footnote{This is more subtle for the metric propagator and more generally when gauge d.o.f.s are involved.}.

This short demonstration appears heuristic for two reasons, which can be rigorously answered. First, $X^<$ is technically a quantum operator, however, the operator acting on it will only leave terms with a time derivative acting on the window after simplification of the linear EOM, meaning that the UV modes will only contribute at the time of crossing, where they are classicalised and can be treated as stochastic numbers with \ylb{the spectrum set by their state at crossing, quantum-originated in the case of standard models of inflation}.
Second, the case of full GR and beyond is more subtle because of the gauge-dependence as pointed out in \cite{Launay24}. The procedure used in the cited work and the present one successfully tackles that, as explained in Sections~\ref{sec:slicingsOfR} and \ref{sec:stoEOMs}.

Note that this work does not look at the validity of the separation of linear quantum vs nonlinear classical scales, and as advocated in \cite{Launay2024bis, Ballesteros26},
assumes that every scenario of inflation and every set of scales selected by $W$ should have a classicality criterion checked using QFT-based arguments in their perturbative phase.
Instead, this work leaves the choice of the window function unspecified and focuses on applying the stochastic ansatz described above to the most general variety of theories, without making any further approximation on the dynamics.
Keeping those notations general will also be fruitful to discuss application beyond the standard case of inflation in Section~\ref{sec:discussion}.

\textit{In the following, we apply the stochastic ansatz to a wide class of \yl{scalar-tensor theories} beyond GR at the same time (i.e. not one by one specifically) by using an EFT description gathering all linear descriptions at once.}

\ylb{\paragraph{Assumptions Summary.} Unless stated otherwise, Sections~\ref{sec:procedure} and \ref{sec:EOMs} assume:
\begin{itemize}[nosep]
  \item a single clock: one scalar degree of freedom, whose background sets the unitary-gauge time slicing;
  \item no matter or radiation sector beyond this scalar and the metric;
  \item linearity of the UV modes on the IR background, as in Eq.~\eqref{eq:actionExpansion};
  \item classicality of the IR modes, which should be checked for each scenario and choice of $W$ \cite{Launay2024bis, Ballesteros26}.
\end{itemize}
They do not assume:
\begin{itemize}[nosep]
  \item a specific theory within the EFT action, its coefficients being left free until Section~\ref{sec:EOMs};
  \item the decoupling limit, mixings with the metric perturbations being kept;
  \item a specific window function $W$;
  \item a Bunch--Davies vacuum: the sources are given by the state of the UV modes at crossing, which need not be quantum (Eq.~\eqref{eq:Qsol} is one example);
  \item the freezing of the UV modes at crossing, their evolution continuing nonlinearly in the IR.
\end{itemize}}

\subsection{\yl{EFT primer} \label{sec:EFToDE}}

Given the linear origin of the sources, the EFT is the ideal framework to describe the dynamics of the UV modes and compute the sources of all included theories simultaneously.
It provides indeed a unified way to write down the most general action for a single scalar degree of freedom coupled to gravity, without having to specify the underlying theory.
Later in Section~\ref{sec:EOMs}, we will specify some of the EFT coefficients to match some known higher-order relativistic theories.
The following subsections first introduce the reader to these tools and notations, mostly taken from \cite{Gleyzes13,Gleyzes14}.

\subsubsection{Symmetries and actions}
\ylb{The effective field theory of a single scalar degree of freedom
spontaneously breaking time diffeomorphisms (the EFT of inflation, EFToI) is a powerful framework to describe the dynamics of the inflaton and its interactions with gravity \cite{Cheung08}}.
It is based on the idea that the dynamics of the scalar field can be described as a low-energy effective theory, where the high-energy degrees of freedom are integrated out. This allows us to capture the essential features without needing to specify its microscopic origin, \yl{and thus without needing the fully nonlinear equations of motion.}
The action of the EFT is written in the unitary gauge as
\begin{equation}
    \begin{aligned}
        {\rm S}  &= \int d^4x \sqrt{-g} \left[ \frac{\Mpl^2}{2} f(t)\,{}^{(4)}R - \Lambda(t) - c(t)g^{00}\right.  \\
       &  \left.  + \frac{M_2^4(t)}{2}(\delta g^{00})^2 - \frac{m_3^3(t)}{2} \delta K \delta g^{00}
       - m_4^2(t) \left(\delta K^2 - \delta K^\mu{}_\nu \delta K^\nu{}_\mu\right) + \frac{\tilde{m}_4^2(t)}{2} R\, \delta g^{00} +\ldots\right],
    \end{aligned}
    \label{eq:EFTaction}
\end{equation}
where $\Mpl$ is the Planck mass, ${}^{(4)}R$ and $R$ are the $4$-dimensional and $3$-dimensional Ricci scalars respectively, and $K_{\mu\nu}$ is the extrinsic curvature. \yl{In this form}, this is the most general action for a single scalar degree of freedom coupled to gravity, up to second order in perturbations and with at most two derivatives in the quadratic action.
The first line is to be understood as the background action, while the second one represents the quadratic expansion around the FLRW metric.
\yl{$f$ is the (possibly time-dependent) dimensionless coefficient renormalising the Planck mass in the gravitational sector, while} $\Lambda$ and $c$ play the role of the scalar potential and kinetic energy respectively, and the diverse time-dependent mass coefficients encode its self and metric interactions.
\yl{Note that this identification of $\Lambda$ and $c$ with a potential and a kinetic energy is only exact once a background trajectory $\phi(t)$ has been chosen, as in eqn.~\eqref{eq:inflationbgGR} below for inflation: off that limit, $\Lambda(t)$ remains a background quantity, while $V(\phi)$ is a field-space function, and the two coincide only along the chosen trajectory.}
In GR, the coefficients above take the values $M_2 = m_3 = m_4 = \tilde{m}_4 = 0$ and $f= 1$.

The least action principle yields the background equations of motion
\begin{align}
    c + \Lambda &= 3\Mpl^2\left(fH^2 + \dot{f}H\right) \equiv \rho_{\yl{\pi}} \,, \\
    \Lambda - c &= \Mpl^2\left(2f\ddot{H} + 3fH^2 + 2\dot{f}H + \ddot{f}\right)  \equiv p_{\yl{\pi}} \,,
\end{align}
where $H \equiv \dot{a}/a$ is the Hubble rate and $\rho_{\pi}$, $p_{\pi}$ are \yl{the boson's} energy density and pressure.
\ylb{In GR inflation with a potential, the background parameters become}
 \begin{subequations}
  \begin{align}
    \Lambda&=\Mpl^2 H^2(3-\varepsilon_1)=V(\phi),\\
     c&=\dot{\phi}^2/2 = \Mpl^2 H^2 \varepsilon_1,
  \end{align}
  \label{eq:inflationbgGR}
 \end{subequations}
where $\varepsilon_1 \equiv -\dot{H}/(H^2)$ is the first Hubble slow-roll parameter, small enough for long enough to ensure a sufficient number of e-folds of inflation and a quasi-scale-invariant power spectrum of perturbations \cite{PlanckInflation}.
\ylb{If we were to work with late-time dark energy models, the EFT would have a different background evolution and include matter coupled to gravity.}
For the sake of generality, we will keep general notations.

At this stage, the action suffers from a broken time diffeomorphism invariance, which can be restored by introducing a scalar field $\pi$ that transforms under time diffeomorphisms. 
This scalar field can be thought of as the Goldstone boson associated with the breaking of time diffeomorphisms.
By performing the Stueckelberg trick $t \to t + \pi(x)$, we can restore the full diffeomorphism invariance of the theory and analyze the dynamics of perturbations gauge-agnostically.
The procedure is rather involving as it also requires the modification of the space foliation used in the definition of the extrinsic curvature.
This was performed in past work, including the specific but pioneering case of inflation \cite{Cheung08}. For our purpose we will only need the entailed linearized equations of motion known from past works.

\ylb{As opposed to late-time dark energy models, it is worth mentioning that the EFToI usually} takes a further step in the approximations by assuming the decoupling limit, which corresponds to neglecting the mixings with gravity and thus the metric perturbations.
This is why a great part of the literature focuses on the interactions of $\pi$ alone but we do not make this approximation here, as we want to capture the full dynamics of the system, including the mixings with gravity, which can be important in some regimes touched by e.g. stochastic inflation.

\yl{
\textit{For the purpose of generality and interpretability beyond the decoupling limit, \ylb{we will use generic dark-energy EFT notations} for the study of inflation. Sections~\ref{sec:beyondinflaton} and \ref{sec:beyondinflation} will discuss applications beyond the inflaton.}
}
\subsubsection{An interpretable parametrisation}
Rarely used in the context of inflation, another parametrisation of the quadratic terms in eqn.~\eqref{eq:EFTaction} is possible and more interpretable.
The so-called $\alpha$-\textit{basis} \cite{BelliniSawicki14} uses five dimensionless and independent functions of time:
\begin{itemize}
  \item $\aK(t)$: \textit{Kineticity}, which quantifies the kinetic energy of the scalar field perturbations
  \item $\aB(t)$: \textit{Braiding}, which describes the mixing between the scalar field and the metric perturbations
  \item $\aT(t)$: \textit{Tensor speed excess}, which measures the deviation of the speed of gravitational waves from the speed of light
  \item $\aM(t)$: \textit{Planck-mass running rate}, which quantifies the rate of change of the effective Planck mass with time
  \item $\aH(t)$: \textit{Horndeski Planck-mass running}, which captures the time variation of the effective Planck mass in Horndeski theories. When set to zero, the action describes all Horndeski quadratic actions.
\end{itemize}
Analogously to the Hubble slow-roll parameters, we also introduce the notation
$$H\alpha_{X}^{(n+1)} \equiv - d_t{\ln \alpha_X^{(n)}},$$ with $\alpha_X^{(0)} = \alpha_X$ for $X = K, B, H, T, M$.

With this basis, the quadratic action reads \cite{Gleyzes14}
\begin{equation}
    \begin{aligned}
    {\rm S}^{(2)} = \int d^3x\, dt\, a^3 \frac{M^2}{2} \Bigg[ \delta K_{ij}\delta K^{ij} - \delta K^2 + (1+\alpha_T)\left(R\frac{\delta\sqrt{h}}{a^3} + \delta_2 R\right) 
+ \alpha_K H^2 \delta N^2  \\
 + 4\alpha_B H\, \delta K\, \delta N + (1+\alpha_H) R\, \delta N \Bigg],    
    \end{aligned}
\end{equation}
where $h_{ij}$ is the spatial metric, $\delta N$ is the perturbation of the lapse function, and the running Planck mass $M$ relates to $\aM$, $\aT$ and $\Mpl$ as
\begin{subequations}
  \begin{align}
  \aM & =\frac{1}{H} \frac{\d}{\d t} \ln M^2, \\
  M^2 & = \frac{\Mpl^2f}{1+\aT},
  \end{align}
\end{subequations}
with other conversions from eqn.~\eqref{eq:EFTaction} to the $\alpha$-basis given in \cite{Gleyzes14}.

Setting $\aB=\aH=\aT=\aM=0$ and $f=1$, gives back GR, while $\aK=2\varepsilon_1$ is the inflationary limit of the kineticity.

Note that one can also go beyond this framework and include \yl{spatial or time derivatives beyond second order}, in the latter case as long as they do not give rise to Ostrogradsky instabilities and ghosts.
We will comment on generalisation to those cases in Section~\ref{sec:discussion}.

\subsection{Linear cosmology of the EFT}
% We work in Fourier space with wavenumber $k$. 
Along the $\pi$ boson perturbation, we write the scalar perturbations of the metric with the following convention
\begin{equation}
    \begin{aligned}
        g_{00} & = -1 - 2\Psi, \\
        g_{0i} & = 2 a^2 B_{,i} \\
              & = 2\alpha_{,i}, \\
        g_{ij} & = a^2 \Big((1 - 2\Phi) \delta_{ij}+2E_{,ij}\Big)\\
               & = a^2(1 - 2\tilde\Phi) \delta_{ij}-2{\rm P}_{ij}\Delta\beta,
    \end{aligned}
\end{equation}
where the secundary notations are those of \cite{Gleyzes13}, and where we defined \yl{the background Laplacian $\Delta = \partial^i\partial_i$, and
the (transverse) projection operator as
\begin{equation}
  {\rm P}_{ij} \equiv {\partial_i \Delta^{-1} \partial_j} - \frac{1}{3}g_{ij} = a^2\Bigg( \frac{k_ik_j}{k^2}- \frac{1}{3}\delta_{ij}\Bigg).
  \label{eq:defPij}
\end{equation}
To convert the on-shell EOM tensors from the notations of \cite{Gleyzes13,Gleyzes14} to ours, we use in the following
\begin{equation}
  {\cal E}^{(1)} = \frac{1}{2}\frac{\delta {\rm S}^{(2)}}{\delta \tilde{\Phi}},
  \qquad \tilde{\cal E}^{(1)} = -\frac{1}{2\Delta}\frac{\delta {\rm S}^{(2)}}{\delta \beta},
  \qquad {\cal E}^{(1)}_{00} = \frac{1}{2}\frac{\delta {\rm S}^{(2)}}{\delta \Psi},
   \qquad {\cal E}^{(1)}_{0i} = -\frac{a^2}{2\Delta}\partial_i\frac{\delta {\rm S}^{(2)}}{\delta \alpha},
\end{equation}
where for spatial metric variations we used
\begin{equation}
  \frac{\delta {}^{(1)}g^{ij}}{\delta \tilde{\Phi}}= 2\bar{g}^{ij}, \qquad \frac{\delta {}^{(1)}g^{ij}}{\delta \beta}= -2{\rm P}^{ij}\Delta.
\end{equation}
}

\subsubsection{Gauge-agnostic equations\label{sec:linEqns}}

The linear equations of motion were derived in \cite{Gleyzes13} in the Newtonian Gauge.
We used those and recovered independently\footnote{
Note that Appendix D of \cite{Gleyzes13} contains typos in the display: 
$H$ factor missing on the $f''$ term of $A_{\pi}$;
the simplification $E_{\pi}$ should yield $-6\dot{H}c$ as opposed to $+6\dot{H}c$; the $6 \Mpl^2 f'(...)$ term of $E_{\psi}$ the factor $2$ should be on $H^2$ not $H'$
} the $\alpha$ basis linear equations, as displayed in \cite{Gleyzes14} in this gauge for $ {\cal E}^{\mu}{}_{\nu}$ (up-down indices linear Einstein equations).
For use in post-inflationary studies, the literature usually showcases either the Newtonian or the synchronous gauge equations. 
This work does not restric to such choice. It is possible to leave the Newtonian gauge for a gauge-agnostic treatment using the transformation
\begin{equation}
    \begin{aligned}
        \Psi & \rightarrow \Psi - \dot{\chi}, \\
        \Phi & \rightarrow \Phi + H\chi, \\
        \pi & \rightarrow \pi - \chi\,,
    \end{aligned}
\end{equation}
where $\chi=-a^2(B-\dot{E})$ rehabilitates arbitrary $B$ and $E$ perturbations of the $g_{0i}$ and $g_{ij}$ metric components.
If one is interested in the linear equations including the boson $\pi$, performing the Stueckelberg trick at this stage is easier than in the EFT action.
This transformation yields the following gauge-agnostic linear equations of motion and their constraints for the \yl{EFT action} in the $\alpha$ basis
\begin{subequations}
  \begin{align}
    {\cal E}_{00}^{(1)} & =0, \label{eq:linEOO}\\
    {\cal E}_{0i}^{(1)} & = \partial_i Q =0,\label{eq:linEOi}\\%-i k_i Q = 0 \\
    {\cal E}_{ij}^{(1)} & = \frac{1}{3}{\cal E}^{(1)}a^2\delta_{ij} + \frac{3}{2}{\rm P}_{ij}\tilde{\cal E}^{(1)} = 0, \label{eq:linIJ}\\
    {\cal E}_\pi^{(1)} & = 0,\label{eq:linpi}
  \end{align}
  \label{eq:linEq}
\end{subequations}
\yl{where we defined trace and traceless parts with the projector ${\rm P}_{ij}$ defined in eqn.~\eqref{eq:defPij},}
and the linear Einstein EOM lhs terms are
\begin{subequations}
\begin{align}
  {\cal E}_{00}^{(1)} &\equiv
  2\Delta\Big(\Phi(1+\aH)+H\chi(1+\aB)+H\pi(\aH-\aB)\Big) \nonumber\\
  & -6H(1+\aB)\dot\Phi+6 H^3\varepsilon_1(1+\aB)\pi +(6\aB-\aK) H\dot\pi+H(-6-12\aB+\aK)\Psi, \label{eq:E00} \\
   Q &\equiv
  4a^2\Delta\Big(-\dot\Phi+H^2\varepsilon_1\pi+\aB H\dot\pi-(1+\aB)H\Psi\Big), \label{eq:Q} \\
  {\cal E}^{(1)} &\equiv -3\tilde{\cal E}^{(1)}+\ddot\Phi-\aB H \ddot\pi+(1+\aB)H\dot\Psi
    +(-3+2\varepsilon_1+\eta-\aM)\varepsilon_1 H^3\pi+(3+\aM)H\dot\Phi \nonumber\\
  &+\bigg((3-\varepsilon_1+\aM)(1+\aB)-\aB^{(1)}\aB\bigg)H^2\Psi
      +\bigg(\varepsilon_1(\aB-1)-\aB(3+\aM-\aB^{(1)})\bigg)H^2\dot\pi, \label{eq:calE}\\
  \tilde{\cal E}^{(1)} &\equiv
  \frac{2}{3}\Delta\Big(\dot\chi+\aH\dot\pi-(1+\aH)\Psi+(1+\aT)\Phi
    +(1+\aM)H\chi+(\aT-\aM)H\pi\Big).
  \label{eq:tildeE}
\end{align}  \label{eq:lineq2}
\end{subequations}
Note that, by orthogonality, ${\cal E}^{(1)}$ and   $\tilde{\cal E}^{(1)}$ are both null on-shell and the latter could be injected in the expression of the former in eqn.~\eqref{eq:calE}.
However, we keep it this way as the coarse-grained theory will not be on-shell, as explained in Section~\ref{sec:ansatz}.

More involving, the linear EOM of $\pi$ is given by
\begin{equation}
\begin{split}
 {\cal E}_\pi^{(1)} = &
  -12\varepsilon_1\Big(\varepsilon_1(1+3\aB)+\aB\big(-3+\eta-\aM+\aB^{(1)}\big)\Big)H^4\pi
  \\&
  -12\big(3\aB+\aB(\aM-\aB^{(1)})-\varepsilon_1(1+\aB)\big)H^2\dot\Phi -12\aB H\ddot\Phi
  \\&
  -2(6\aB-\aK)H^2\dot\Psi -2\aK\Big(3+\aM-2\varepsilon_1-\aK^{(1)}\Big)H^3\dot\pi -2\aK H^2\ddot\pi
  \\&
  -2\Big((6\aB-\aK)(3+\aM)-2\varepsilon_1(3+9\aB-\aK)+\aK\aK^{(1)}-6\aB\aB^{(1)}\Big)H^3\Psi
  \\&
  +4\Delta\Bigg(\Big(\aB(1+\aM-\aB^{(1)})-\varepsilon_1(1+\aB)\Big)H^2\chi + \aH \dot\Phi \\&\qquad
   +\aB H\dot\chi+(\aH-\aB)H\Psi+ \big(\aM+\aH(1+\aM-\aH^{(1)})-\aT\big)H\Phi \\&\quad
  +\Big(\varepsilon_1(1+\aB-\aH)+\aM
    -\aB(1+\aM-\aB^{(1)})+\aH(1+\aM-\aH^{(1)})-\aT\Big)H^2\pi \Bigg).
\end{split}
  \label{eq:Epi}
\end{equation}
We are not interested in the hydrodynamical equations, and so do not rewrite the $\dot{\rho}$ evolution equation found in \cite{Gleyzes13,Gleyzes14}.
A related discussion on potential extensions is lead in Section~\ref{sec:discussion}. The same remark is valid for vector modes.
Spin-2 linear equations are easier to write than the scalar ones as the gauge-invariance of the tensorial perturbations mean that the linear contribution to the Einstein equations is proportional to the polarised Mukhanov-Sasaki equations.
We thus focus on scalars until Section~\ref{sec:stochgrav} without loss of generality.

\subsubsection{Curvature perturbation dynamics}
Despite the complexity of eqn.~\eqref{eq:linEq} and \eqref{eq:lineq2}, a single gauge-invariant perturbation can summarise the linear scalar dynamics,
as we recall below using the curvature perturbation on comoving hypersurfaces
\begin{equation}
{\cal R} \equiv \Phi+{H}\pi.
\end{equation} % note: it's a +sign because zeta is - R.
\paragraph{Mukhanov-Sasaki equation.}
Define the field $z$ as 
\begin{equation}
  z^2 \equiv\frac{ a^3 M^2\abar}{(1+\aB)^2} = \frac{a^3 \Mpl^2 f\,\abar}{(1+\aT)(1+\aB)^2}\,,
\end{equation}
and the nonlinear combination $\abar \equiv \aK + 6\aB^2$. The logarithmic cosmic time derivative is
\begin{equation}
 2\dot{\overbrace{\ln{z}}} = H(3+\aM)
  - H\frac{2(6\aB-\aK)\aB\aB^{(1)} + (1+\aB)\aK\aK^{(1)}}{(1+\aB)\abar}.
\end{equation}
A cosmologist would notably find back the $-H\varepsilon_2$ term of the GR Mukhanov-Sasaki equation in $\aK^{(1)}$. 
Define now the sound speed
\begin{equation}
  c_s^2 \equiv -2\frac{(1+\aB)^2}{\bar\alpha}\left[
    1+\aT
    - \frac{1+\aH}{1+\aB}\left(
      1 + \aM +\varepsilon_1
      %+ \frac{1}{H}\frac{d}{dt}\ln\frac{1+\aH}{1+\aB}
      +\frac{\aB\aB^{(1)}}{1+\aB}-\frac{\aH\aH^{(1)}}{1+\aH}
    \right)
  \right].
\end{equation}
The Mukhanov-Sasaki equation \yl{\cite{Mukhanov1981,Sasaki83}} can be found by rewriting the \yl{EFT action} or the linear equations in the unitary gauge, giving \cite{Gleyzes13}
\begin{equation}
  \ddot{\R} +  2\dot{\overbrace{\ln{z}}}\,\dot{\R} + \frac{c_s^2 k^2}{a^2}\,\R = 0.
  \label{eq:MS}
\end{equation}
For the case of inflation, and similarly to GR, these modes could be given Bunch-Davies initial conditions in the sub-Hubble regime.
\yl{This is only one initial-condition choice, and the present formalism remains valid as long as $\R$ remains perturbative; it does not require the modes to freeze, since their evolution simply continues in the IR nonlinearly.}

\begin{figure}
  \centering
  \includegraphics[width=0.65\textwidth]{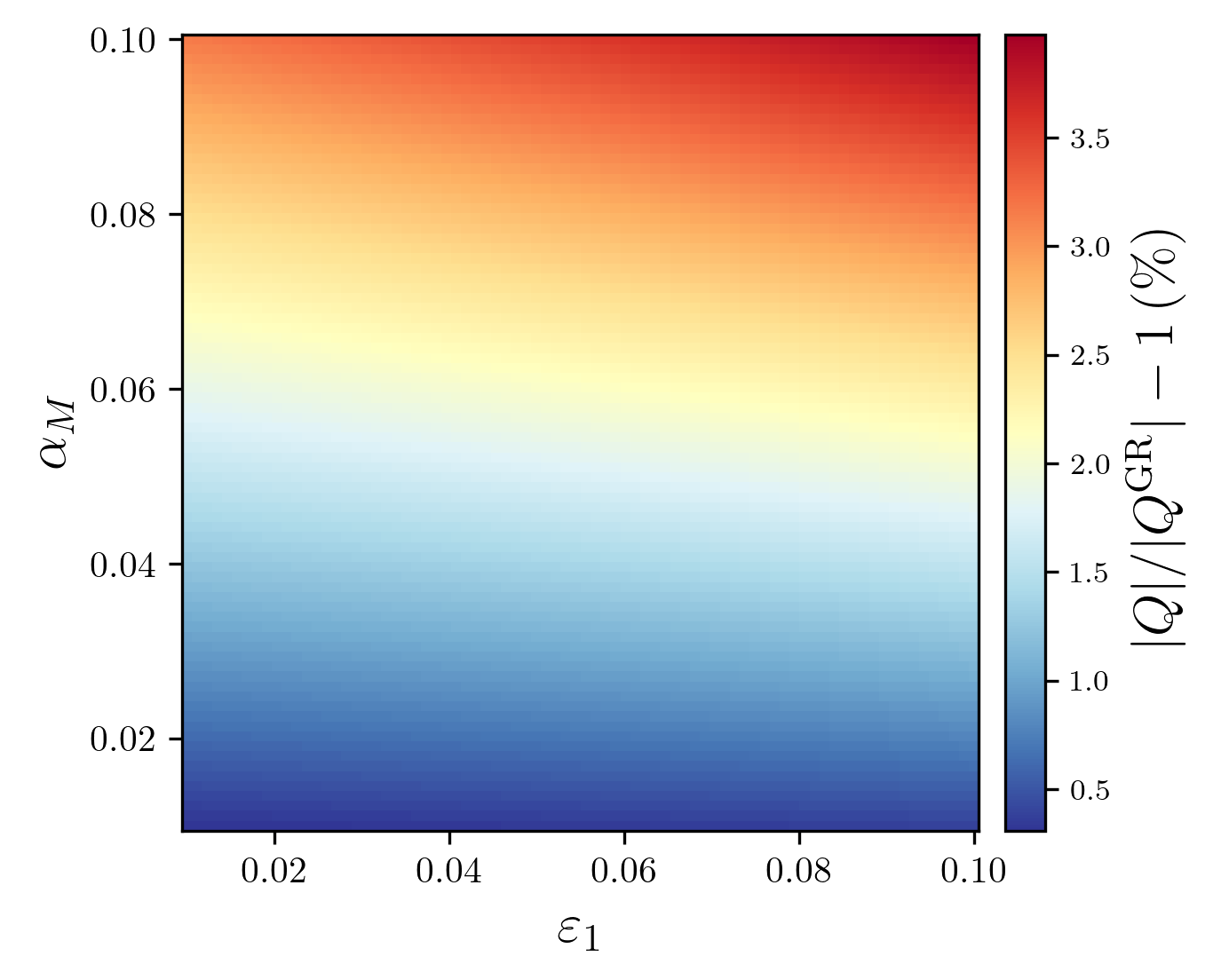}
  \caption{Relative departure of the inflationary Mukhanov-Sasaki variable from its GR value at horizon crossing ($-kc_s\tau=1$) as a function of the Planck mass running $\aM$ and the first Hubble slow-roll parameter $\varepsilon_1$. All Hubble and EFT parameters are assumed slowly-varying.}
  \label{fig:proxyQ}
\end{figure}

The value of a ${\cal R}$ mode at (chosen) crossing to the IR has a direct impact on the amplitude of the stochastic sources.
Let us look at its value in the case of inflation. By defining $Q = z{\cal R}$, we can rewrite eqn.~\eqref{eq:MS} in conformal time $\tau$ as
\begin{equation}
  Q_k''+\bigg(k^2c_s^2-\frac{z''}{z}\bigg)Q_k=0,
\end{equation}
and assume slowly-varying Hubble and EFT rolling parameters at crossing ($-kc_s\tau=1$), so that the solution is analytically known as
\begin{equation}
  Q_k = A_k(-kc_s\tau)^{1/2}H_{\nu_s}^{(1)}(-kc_s\tau)+ B_k(-kc_s\tau)^{1/2}H_{\nu_s}^{(2)}(-kc_s\tau),
  \label{eq:Qsol}
\end{equation}
where $\nu_s^2 = \frac{1}{4}+\frac{2-\varepsilon_1+3\alpha_M/2}{(1-\varepsilon_1)^2}$.
\ylb{Bunch--Davies initial conditions in the deep sub-Hubble regime, $Q_k \to e^{-ic_sk\tau}/\sqrt{2c_sk}$ for $-kc_s\tau\to\infty$, give $B_k=0$ and $A_k = \frac{\sqrt{\pi}}{2}e^{i(\nu_s+1/2)\pi/2}/\sqrt{c_sk}$.
This is where the quantum input enters: quantisation imposes $|A_k|^2-|B_k|^2=\pi/(4c_sk)$, which fixes the amplitude of the sources. Other initial states, e.g. $\alpha$-vacua ($B_k\neq0$) \cite{Mottola85, Bruce85} or classical initial conditions, would only change $A_k$ and $B_k$; the rest of the procedure does not require quantum initial conditions but non-zero Fourier modes for ${\cal R}$.}
For the Bunch-Davies example, we provide a visual comparison of $Q(-kc_s\tau=1)$ with and without running of the Planck mass in Figure~\ref{fig:proxyQ}.
A few percents deviation can be reached with a reasonably perturbative $\aM$ value, translating in a similar difference per efold in the amplitude of the stochastics we are about to compute.

\subsubsection{Perturbations as functionals of ${\cal R}$ \label{sec:slicingsOfR}}

Following the procedure developed in \cite{Launay24} before performing the coarse-graining, we express the gauge invariant combinations in terms of $\R$. 
These are easily obtained in the comoving gauge where $\pi = E = 0$ and where $\Phi = \R$ by definition. 

In the comoving gauge the momentum constraint eqn.~\eqref{eq:linEOi} reduces to solving
$\dot{\R} + \Psi H(1+\aB) = 0$, giving
\begin{equation}
    {\Psi^{\rm co} = -\frac{1}{H(1+\aB)}\dot{\R}\,.}
\end{equation}
Substituting the above into the Hamiltonian constraint eqn.~\eqref{eq:linEOO} yields
\begin{equation}
  {\chi^{\rm co} = -\frac{1}{2(1+\aB)^2}\bigg(
    \frac{2(1+\aB)(1+\aH)}{H}\R
    + \frac{\, a^2\abar}{k^2}\dot{\R}
  \bigg),}
\end{equation}
which here equates $-a^2B^{\rm co}$. The energy density perturbation is found too as $\delta\rho^{\rm co} = -2c \Psi^{\rm co}[\R]$.
We can now evaluate the common gauge invariants used in cosmology, as functionals of $\R$
\begin{subequations}
\begin{align}
  \Phi_B     &\equiv \Phi + H\chi = \R + H\chi^{\rm co}[\R]\,, \\
  \Psi_B     &\equiv \Psi - \dot\chi = \Psi^{\rm co}[\R] - \dot\chi^{\rm co}[\R]\,, \\
  \delta\phi^{\rm gi} &\equiv \frac{\pi}{\sqrt{2c}} - \sqrt{2c}\chi = -\sqrt{2c}\chi^{\rm co}[\R]\,, \\
  \zeta^{\rm gi}  &\equiv -\Phi + \frac{1}{3}\frac{\delta\rho}{\rho_{\yl{\pi}}+p_{\yl{\pi}}} = -\R + \frac{1}{3}\frac{\delta\rho^{\rm co}[\R]}{\rho_{\pi}+p_{\pi}}\,.
\end{align}\label{eq:GIs}
\end{subequations}
where $\rho_{\pi}+p_{\pi} = 2c + M^2(\ddot f - H\dot f)$ is the effective \yl{boson} energy density plus pressure, and where we referred to the derived comoving gauge functionals we just derived explicitly.
As shown in our previous work in GR for inflation \cite{Launay24}, knowing these as functionals of $\R$ is particularly useful to get all gauges expressed as functionals of $\R$.
In particular in the relevant gauges, this provides physically-motivated cosmological initial conditions for numerical relativity in modified gravity, without relying on arbitrary initial perturbations and constraint solvers \yl{\cite{brady_inflaton_2025}}.

Finally, one can compute the gravitational slip as
\begin{subequations}
  \begin{align}
  \Phi_B - \Psi_B & = \R + H\chi^{\rm co}[\R] - \Psi^{\rm co}[\R] + \dot\chi^{\rm co}[\R]\,. \\
& =  \Bigg( 1 -\frac{1+\aH}{1+\aB}\Big(1+\varepsilon_1+
      \frac{\aB\aB^{(1)}}{1+\aB}-\frac{\aH\aH^{(1)}}{1+\aH}
    \Big) \Bigg) \R \nonumber\\
& +  \Bigg( \frac{ \abar\Big( \aM-2\dot{\overbrace{\ln{z}}} \Big)}{2 (1+\aB)^2}\Big(\frac{k}{aH}\Big)^{-2} - \frac{ \aH}{1+\aB} \Bigg)\frac{\dot \R}{H} - \frac{\abar}{2 (1+\alpha_B)^2} \Big(\frac{k}{aH}\Big)^{-2} \frac{\ddot \R}{H^2}\,\\
& =  \Big(\frac{1+\alpha_H}{1+\alpha_B}\alpha_M-\alpha_T\Big)\R +  \Bigg(
\frac{ \abar\aM}{2 (1+\aB)^2}\Big(\frac{k}{aH}\Big)^{-2}- \frac{\alpha_H}{(1+\alpha_B)}
\Bigg)\frac{\dot \R}{H}\, .
\end{align}
\end{subequations}
In the second line, we injected the Mukhanov-Sasaki equation, eqn.~\eqref{eq:MS}. In GR, this becomes clearly $0$ as expected, but not in the general case.
In this study, it is important not to inject the Mukhanov-Sasaki equation in the expression of the gravitational slip or any EOMs before coarse-graining.
In fact, the coarse-grained gravitational slip is non-zero, even in GR \cite{Launay24}, \yl{which is why it is important to keep track of both expressions above.}

\subsection{Sources to all EOMs \label{sec:stoEOMs}}
Now that we have identified the true underlying dynamical field ($\cal R$) in the EOMs, we have everything we need to proceed with the coarse-graining of the linear theory as described in Section~\ref{sec:ansatz}.

\subsubsection{Scalar sector}
As explained in Section~\ref{sec:ansatz}, the coarse-graining procedure consists in seeing the change of the linear equations of motion of a field $X$ under the transformation
\begin{equation}
  X(t,\mathbf{k}) \;\mapsto\; X(t,\mathbf{k})\,W(t,\mathbf{k})\,,
  \label{eq:coarsegrainmap}
\end{equation}
where $W$ is a window function, before promoting the result to be the spectrum of the stochastic source to the nonlinear EOMs.
However in full general relativity, and despite having a single scalar degree of freedom, multiple scalar perturbations are at play with no clear prescription for their coarse-graining as a group.
Moreover, coarse-graining should not be gauge-dependent, and thus should be performed on gauge-invariant combinations of the fields, as we advocated in \cite{Launay24}.
This is why using the coarse-graining $X = {\cal R}$ is the most natural choice, in particular after having expressed all the gauge-invariant combinations as functionals of $\R$ in the previous section.

In fact, eqns.~\eqref{eq:GIs}, allow us to write any field in any gauge as a functional of $\R$. 
If one picks a gauge-dependent perturbation without specifying its value in a given gauge, say $\pi$, then all other gauge-dependent quantities can be expressed as functionals of $\pi$ and $\R$
\begin{subequations}
  \begin{align}
    \chi[\R,\pi]  & = \chi^{\rm co}[\R]+\frac{\pi}{2c},\\
    \Phi[\R,\pi]  & = \R + H\chi^{\rm co}[\R] - H\chi[\R,\pi],\\
    \Psi[\R,\pi]  & = \Psi^{\rm co}[\R]+\dot{\chi}[\R,\pi]-\dot{\chi}^{\rm co}[\R] .
  \end{align}
\end{subequations}
Substituting these three functionals in the linear equations of motion, eqn.~\eqref{eq:linEq} and eqn.~\eqref{eq:linpi}, \yl{will eliminate $\pi$ and give the Mukhanov-Sasaki equations} up to different factors, except for the constraints which will be automatically satisfied by definition and as non-dynamical equations.
As we are about to see, substituting the coarse-grained functionals
\begin{equation}
  \chi^>\equiv\chi[W\R,\pi^>], \qquad \Phi^>\equiv\Phi[W\R,\pi^>], \qquad \Psi^>\equiv\Psi[W\R,\pi^>],
  \label{eq:CGfuncs}
\end{equation}
\yl{where we have applied the Fourier coarse-graining defined in eqn.~\eqref{eq:coarsegrainmap} to $X=\R$,} however, might not give these results, as we are about to see \yl{beyond GR}.

\paragraph{Mukhanov-Sasaki equation.}
Even as a linear EOM, the Mukhanov-Sasaki equation itself can be coarse-grained and will appear useful in the simplification of the sources to the full EOMs.
In Fourier space, ${\cal R}^>(t,\mathbf{k}) = W(t,\mathbf{k}){\cal R}(t,\mathbf{k})$ satisfies the following equation
\begin{equation}
  \ddot{\overbrace{\R^>}} +  2\dot{\overbrace{\ln{z}}}\,\dot{\overbrace{\R^>}}+ \frac{c_s^2 k^2}{a^2}\,\R^> = \Src_{\cal R}^{\rm EFT},
\end{equation}
after simplifying using the MS equation satisfied by ${\cal R}^>$ so that the source Fourier amplitude $ \Src_{\cal R}^{\rm EFT}$ takes the form
\begin{align}
  \Src_{\cal R}^{\rm EFT} &=  \R\ddot W + \big(2\dot \R+2\dot{\overbrace{\ln{z}}}\,\R\big)\dot W\,. \label{eq:SR}
\end{align}
The result in canonical inflation GR of \cite{Launay24}, noted $\Src_{\cal R}^{\rm GR}$, is recovered in this limit where $ 2\dot{\overbrace{\ln{z}}} = H(3-\varepsilon_2)$ and $c_s=1$. This is particularly visible when writing
\begin{equation}
  \Src_{\cal R}^{\rm EFT} = \Src_{\cal R}^{\rm GR}+\gamma\dot W\R,
\end{equation}
with $\gamma$ a function of the $\alpha$ parameters that vanishes in the GR limit and defined as

\begin{equation}
  \gamma \equiv 2\dot{\overbrace{\ln{z}}}-H(3-\varepsilon_2) =  H\Bigg(\varepsilon_2+\aM -\frac{ 2(6\aB-\aK)\aB\aB^{(1)}}{\abar(1+\aB)}-\frac{\aK\aK^{(1)}}{\abar}\Bigg) \,.
\end{equation}

\paragraph{Constraint IR equations.\label{sec:cons}}
We now plug the coarse-grained and gauge-agnostic perturbation functionals \yl{$\chi^>,\Phi^>,\Psi^>$ defined in eqn.~\eqref{eq:CGfuncs}}, in the Hamiltonian and momentum constraints, eqn.~\eqref{eq:linEOO} and eqn.~\eqref{eq:linEOi}, and call $\Src_{\cal H}$ and $\Src_{\cal M}$ the results in Fourier space, after simplifying using the MS equation satisfied by $\R$.
We find that these source spectra \yl{vanish identically}
\begin{equation}
  \Src_{\cal H} = 0\,,\qquad \Src_{\cal M} = 0\,.
\end{equation}
As shown in previous work \cite{Launay24, Launay2025}, this is the benefit of using the gauge-agnostic coarse-graining.
We should stress again, that no previous work in the stochastic inflation literature, apart from \cite{Launay24} and the present one, showed explicit satisfaction of the coarse-grained constraints, which is rather essential to input a physically-balanced stochastic kick at each time step.
At best, the local Friedmann equation corresponding to the Hamiltonian in the separate universe approximation (0th order gradient expansion) would be enforced in most works, and the momentum equation has been used to prepare initial data at next order in the gradient expansion \cite{prokopec_ensuremathdeltan_2021}.
Some works, including in the context of higher curvature corrections, have found a non-zero source to the constraints (see eqns. (61) and (64) of \cite{aldabergenov_towards_2025}). 
We believe and have seen in our own previous attempts that this might be a consequence of coarse-graining in a specific gauge inconsistently (flat gauge in the mentioned work) or simplifying with the MS equation before coarse-graining.

\paragraph{Source to IR scalar.} \yl{Moving beyond the constraints and the MS equation yields much more complexity at first sight: after substituting the functionals of $W\R$ and $\pi^>$ and simplifying $\pi^>$ and the MS equation in $\R$, we get the source Fourier amplitude for eqn.~\eqref{eq:linpi} to be}
\begin{multline}
  \Src_\pi^{\rm EFT}
  = -\frac{2 H}{(1+\aB)^3}\Bigg[
    \R\ddot W\,(1+\aB)\abar
    + \R\dot W\,H(1+\aB)\abar(3+\aM) \\
    + \dot W\Bigl(2\dot \R\,(1+\aB)\abar
      -H \R\bigl[2(6\aB-\aK)\aB\aB^{(1)} + (1+\aB)\aK\aK^{(1)}\bigr]\Bigr)
  \Bigg]\,.
\end{multline}
However, this compactifies to
\begin{equation}
  \Src_\pi^{\rm EFT}
  = -\frac{2 H \abar}{(1+\aB)^2} \Src_{\cal R}^{\rm EFT},
\end{equation}
after identifying the source Fourier amplitude of the MS equation found in eqn.~\eqref{eq:SR}
In the GR limit, this \yl{reads}
\begin{equation}
  \Src_\pi^{\rm GR}
  = -4 H \varepsilon_1 \Src_{\cal R}^{\rm GR}.
\end{equation}
We recover the \yl{Arnowitt-Deser-Misner (ADM)} field equation source Fourier amplitude of \cite{Launay24} using $\delta\phi = \sqrt{2\varepsilon_1}H\Mpl\, \pi$ at linear order\footnote{The minus sign comes from having $-\ddot \pi$ in the least action principle as opposed to the ADM equation.}
\begin{equation}
  \Src_\Pi^{\rm GR}
  = \sqrt{2\varepsilon_1} M_{Pl} \Src_{\cal R}^{\rm GR}.
\end{equation}
Beyond GR, the appropriate field redefinition to recover the same form is $\delta\phi = \sqrt{2c}\, \pi = \dot\phi_b\,\pi$, which can be applied to find the source to ${\cal E}_{\phi}$ as
\begin{equation}
  \Src_\phi^{\rm EFT}
  = -\frac{2 \dot\phi_b H \abar}{(1+\aB)^2} \Src_{\cal R}^{\rm EFT}.
\end{equation}

\paragraph{Source to IR Geometry.} With the same method, the sources to $\cE_{ij}$ are, as observed in the GR case,
proportional to the MS source too and factorise as
\begin{align}
  \Src_{\cal E}^{\rm EFT}
  & = -\frac{\abar}{(1+\aB)^2}\Src_{\cal R}^{\rm EFT},\\
    \Src_{\tilde{\cal E}}^{\rm EFT}
  & =  \frac{\abar}{3(1+\aB)^2}\Src_{\cal R}^{\rm EFT},
\end{align}
\yl{which combine, analogously to the on-shell trace-traceless split of eqn.~\eqref{eq:linIJ}, into the spatial source
\begin{equation}
  \Src_{{\cal E}_{ij}}^{\rm EFT} = \frac{1}{3}\Src_{\cal E}^{\rm EFT}\,a^2\delta_{ij} + \frac{3}{2}{\rm P}_{ij}\,\Src_{\tilde{\cal E}}^{\rm EFT}\,,
  \label{eq:EijTraceSplit}
\end{equation}
}meaning, after substituting the two expressions above and using $\tilde{\rm P}_{ij}$ defined below, that the total source to the Einstein tensor is
\begin{align}
  \Src_{{\cal E}_{ij}}^{\rm EFT} =\frac{\abar}{2(1+\aB)^2} \tilde{\rm P}_{ij}\Src_{\cal R}^{\rm EFT},
\end{align}
where we defined the operator
$$\tilde{\rm P}_{ij} = {\partial_i \Delta^{-1} \partial_j}- \bar{g}_{ij} = a^2\bigg( \frac{k_ik_j}{k^2}- \delta_{ij}\bigg).$$
We thus recover elegantly \cite{Launay24} in the GR limit.

\subsubsection{Stochastic gravitons\label{sec:stochgrav}} 
Linearising in the traceless and divergence-free part of the spatial metric as $g_{ij} = a^2(\delta_{ij} + h_{ij})$, the gravitonic \yl{EFT action} is \cite{Gleyzes13,Gleyzes14}
\begin{equation}
  {\rm S}_h = \frac{1}{2}\int d^4x \frac{a^3 M^2}{4}\left[
    \dot h_{ij}^2 - \frac{c_T^2}{a^2}(\partial_k h_{ij})^2
  \right],
\end{equation}
where the tensor speed can have an excess to celerity $c_T^2 = 1+\aT$.
Hence, using the same approach as for the scalar case and using $z_h^2=\frac{\Mpl^2a^3 f}{4(1+\aT)}$, the effective gravitonic Mukhanov-Sasaki equation writes for the two Fourier IR polarisation modes $h^>$
\begin{equation}
  \ddot{\overbrace{h^>_k}} +  2\dot{\overbrace{\ln{z_h}}}\,\dot{\overbrace{h^>_k}} + \frac{c_T^2 k^2}{a^2}h_k = \Src_{h}^{\rm EFT},
\end{equation}
where the source Fourier amplitude $ \Src_{h}^{\rm EFT}$ takes the form
\begin{align}
  \Src_{h}^{\rm EFT} &=  h\ddot W + \big(2\dot h+2\dot{\overbrace{\ln{z_h}}}\,h\big)\dot W.
\end{align}
The GR result of \cite{Launay24}, noted $\Src_{h}^{\rm GR}$, is recovered in the GR limit where $ 2\dot{\overbrace{\ln{z_h}}} = 3H$ and $c_T=1$. This is particularly visible when writing
\begin{equation}
  \Src_{h}^{\rm EFT} = \Src_{h}^{\rm GR}+\gamma_h\dot W h,
\end{equation}
with $\gamma_h$ a function of the $\alpha$ parameters that vanishes in the GR limit and defined as
\begin{equation}
  \gamma_h \equiv 2\dot{\overbrace{\ln{z_h}}}-3H = \frac{\d}{\d t} \ln \Big(\frac{f}{1+\aT}\Big)\,.
\end{equation}
 At linear order, the stochastic gravitons only contribute to the $\cE_{ij}$ equation in its traceless-transverse part.
This $\Src_{h}^{\rm EFT}$ traceless-transverse source is thus to be added to the r.h.s. of the ${\cal E}_{ij}$ equation only, with a $\frac{a^2}{2}$ factor, summed over the two polarisation modes $s=+,\times$.

\subsubsection{Stochastic scalar-tensor equations \label{sec:mainresult}}
  We now collect the previous findings in a shared formalism.
  Within the scope of validity of the stochastic ansatz, the IR relativistic equations of motion of a scalar-tensor theory with action ${\rm S}$ \yl{and matching EFT coefficients $\{\alpha\}$ write
  \begin{subequations}
    \begin{align}
       {\cal E}_{\phi}[g^>, \pi^>] = & - \frac{2\dot\phi_bH\abar}{(1+\aB)^2}  {\cal F}^{-1}\Big\{\Src_{\cal R}^{\rm EFT}\boldsymbol{\xi}^{(0)}_{\mathbf{k}}\Big\}\\
      {\cal E}_{\mu\nu}[g^>, \pi^>]
          = & \frac{\abar}{2(1+\aB)^2}  {\cal F}^{-1}\Bigg\{
          \begin{pmatrix}
            0 & \boldsymbol{0} \\
            \boldsymbol{0} & \tilde{\rm P}
          \end{pmatrix} \Src_{\cal R}^{\rm EFT}\boldsymbol{\xi}^{(0)}_{\mathbf{k}}\Bigg\}\nonumber\\
          & + \frac{1}{2}a^2\displaystyle\sum_{s=+,\times}
             {\cal F}^{-1}\Bigg\{
            \begin{pmatrix}
            0 & \boldsymbol{0} \\
            \boldsymbol{0} & \varepsilon^{(s)}(\mathbf{k})
          \end{pmatrix} \Src_h^{\rm EFT}\boldsymbol{\xi}^{(s)}_{\mathbf{k}}\Bigg\},
    \end{align}\label{eq:main}
  \end{subequations}
  }where $\forall s=0,+,\times$,
  \begin{equation}
    \langle \boldsymbol{\xi_{\mathbf{k}_1}}^{s_1}\boldsymbol{\xi_{\mathbf{k}_2}}^{s_2*} \rangle_{\mathbb{P}}  =  \delta^{(3)}(\mathbf{k}_1-\mathbf{k}_2)\delta^{s_1s_2}.
  \end{equation}

  % recovering the GR limit when $\abar = \aK = 2\varepsilon_1$, $\aB=\aM=\aT=\aH=0$. 
  It is important to understand that the spin-$0$ stochastic grid $\boldsymbol{\xi}^{(0)}_{\mathbf{k}}$ sees the same realisation in both the scalar field and geometry equations to ensure the displayed satisfaction of the constraints.
After coarse-graining, the original lhs of the (dynamical) EOMs have thus acquired the following power spectra due to their sources in the IR
\begin{subequations}
\begin{align}
    \langle {\cal E}_{\pi}(\mathbf{k}_1) {\cal E}_{\pi}^*(\mathbf{k}_2) \rangle_{\mathbb{P}} & = \frac{4H^2\abar^2}{(1+\aB)^4}\delta^{(3)}(\mathbf{k}_1-\mathbf{k}_2)|{\cal S}_{\cal R}^{\rm EFT}(k_1)|^2\\
    \langle {\cal E}_{ij}(\mathbf{k}_1) {\cal E}_{kl}^*(\mathbf{k}_2) \rangle_{\mathbb{P}} & = \frac{ \abar^2}{4(1+\aB)^4}\delta^{(3)}(\mathbf{k}_1-\mathbf{k}_2)\tilde{\rm P}_{ij}\tilde{\rm P}_{kl} |{\cal S}_{\cal R}^{\rm EFT}(k_1)|^2 \notag\\
                                                                                      & +\frac{1}{4}a^4\delta^{(3)}(\mathbf{k}_1-\mathbf{k}_2)\displaystyle\sum_{s=+,\times}\varepsilon^{s}_{ij}(\mathbf{k}_1)\varepsilon^{s}_{kl}(\mathbf{k}_1)|{\cal S}_h^{\rm EFT}(k_1)|^2 \notag \\
                                & = \frac{ \abar^2}{4(1+\aB)^4}\delta^{(3)}(\mathbf{k}_1-\mathbf{k}_2)\tilde{\rm P}_{ij}\tilde{\rm P}_{kl} |{\cal S}_{\cal R}^{\rm EFT}(k_1)|^2\notag\\
                                & + \frac{1}{4}\delta^{(3)}(\mathbf{k}_1-\mathbf{k}_2) \left(\tilde{\rm P}_{ik}\tilde{\rm P}_{jl}+\tilde{\rm P}_{il}\tilde{\rm P}_{jk}-\tilde{\rm P}_{ij}\tilde{\rm P}_{kl} \right)|{\cal S}_h^{\rm EFT}(k_1)|^2.\
\end{align}
\end{subequations}

In the following, we condense the notations of eqn.~\eqref{eq:main} by writing the spin-0 and spin-2 sources as
\begin{equation}
  \begin{aligned}
     \boldsymbol{{\cal S}}^{(0)} & = {\cal F}^{-1}\{ \Src_{{\cal R}}^{\rm EFT}\boldsymbol{\xi}^{(0)}_{\mathbf{k}}\}, \\
     \boldsymbol{{\cal S}}_{\mu\nu}^{(2)} & = \frac{1}{2}a^2\displaystyle\sum_{s=+,\times} {\cal F}^{-1}\{\varepsilon^{(s)}_{\mu\nu}(\mathbf{k}) \Src_h^{\rm EFT}\boldsymbol{\xi}^{(s)}_{\mathbf{k}}\},
  \end{aligned}
\end{equation}
together with the convention $\tilde{\rm P}_{00} = \tilde{\rm P}_{0i}= \tilde{\rm P}_{i0} = 0$, and similarly for the polarisation tensors.

Furthermore, we also drop the superscript $^>$ notation of IR fields and keep it implicit.

\section{Examples in motion\label{sec:EOMs}}

To illustrate the potential and generality of what has been shown in the previous section, we pick here historical extensions to General Relativity and find their associated stochastic EOMs for inflation, simply knowing their quadratic action.
For comparison, only a few examples have made it to the stochastic framework to our knowledge \cite{garcia-bellido_jordan-brans-dicke_1994, aldabergenov_towards_2025}, all in the long wavelength limit. We leave the study of this regime and its validity to future work.

A great deal of these theories can be mapped to the Einstein frame (some of which in the next subsection below), and so, with a singular scalar at stake, can be studied within stochastic inflation in GR \cite{Launay24}, using simply a modified potential.
However, this is not the case for an arbitrary scalar-tensor theory.
For the theories which can be recasted to GR, it is still important to master the stochastic formalism as, generically, going beyond single field phenomenology would not make that possible.

Attached to these examples, we provide Tables~\ref{tab:bgMatch} and \ref{tab:alphaMatch}, which summarise the background and linear matching to the \yl{EFT action} parameters, which are the only ingredients needed to find the stochastic EOMs of a theory within our formalism.

\subsection{Stochastic Einstein-Gauss-Bonnet dynamics\label{sec:EGB}}
\textit{The results of this section were obtained using both the procedure specific to the EGB equations and the procedure matching to the \yl{EFT action} sources. }

The Einstein-Gauss-Bonnet (EGB) theory is a higher-order scalar-tensor theory that includes a coupling between a scalar field and the Gauss-Bonnet invariant.
It is a well-studied model \yl{and} has been shown to have interesting phenomenological implications, including in the context of inflation \cite{fernandes_4d_2022,brady_inflaton_2025},
notably in the bias to the effective potential's slope enabling GB-induced ultra slow-roll and so its own Primordial Black Hole production channel \cite{Shinsuke21,aldabergenov_towards_2025}.

Let us write the EGB action as
\begin{equation}
  \Src_{\rm EGB} = \int d^4x\sqrt{-g}\left[
    \frac{\Mpl^2}{2}R
    - \frac{1}{2}\nabla^\mu\phi\nabla_\mu\phi - V(\phi)
    -\frac{1}{8} \lambda(\phi)\,\mathcal{G}
  \right],
\end{equation}
where $\mathcal{G} = R^2 - 4R_{\mu\nu}R^{\mu\nu} + R_{\mu\nu\rho\sigma}R^{\mu\nu\rho\sigma}$
is the Gauss-Bonnet density and $\lambda(\phi)$ is a dimensionless coupling function.
Let us \yl{next define} the GB-induced energy-momentum tensor, to be compared to the GR one
 \begin{equation} 
\begin{aligned}
T^{\mathrm{GB}}_{\mu\nu} & =
 R_{\mu\rho\nu\sigma} \, \nabla^\rho \nabla^\sigma \lambda
- (\square \lambda) R_{\mu\nu} + 2\, \nabla_\rho \nabla_{(\mu} \lambda \, R^\rho_{\ \nu)} \\
&- \frac{1}{2} (\nabla_\mu \nabla_\nu \lambda)\, R - \frac{1}{2} \left(
2\, \nabla_\rho \nabla_\sigma \lambda \, R^{\rho\sigma}
- (\square \lambda)\, R
\right) g_{\mu\nu}, \\
T_{\mu\nu}^{\phi} & =
\nabla_\mu \phi \, \nabla_\nu \phi - \left(V(\phi)+ \frac{1}{2} \nabla^\rho \phi \, \nabla_\rho \phi \right)g_{\mu\nu}, 
\end{aligned}
\end{equation}
with which we can write the nonlinear EGB equations of motion as
\begin{equation}
  \begin{aligned}
  G_{\mu\nu} - \frac{1}{\Mpl^2}\Big(T^{GB}_{\mu\nu} + T^{\phi}_{\mu\nu}\Big) &=0, \\
    \nabla_\mu\nabla^\mu\phi -\frac{1}{8}\lambda_{,\phi}(\phi){\cal G} - V_{,\phi}(\phi) & = 0. \\
  \end{aligned}
\end{equation}
The 00, ii, and inflaton LHS background equations are respectively given by 
\begin{align}
  3 H^2(1-w) - \frac{1}{\Mpl^2}\Big(\frac{1}{2}\dot\phi^2_b + V(\phi_b)\Big)& =0\,, \\
 a^2 H^2\Big(-3+2\varepsilon_1+w(2- \varepsilon_1+\sigma)\Big)   - \frac{a^2}{\Mpl^2}\Big(\frac{1}{2}\dot\phi^2_b - V(\phi_b)\Big) & = 0,\\
  -\ddot\phi_b -3H\dot\phi_b -V_{,\phi}(\phi_b) - 3w(1-\varepsilon_1)\frac{H^3\Mpl^2}{\dot\phi_b}&=0,
\end{align}
where we have defined the dimensionless coupling parameters
\begin{equation}
  \begin{aligned}
    w & \equiv H \Mpl^{-2} \dot \lambda\,,\\
    \sigma & \equiv H^{-1}d_t\ln w\,.
  \end{aligned}
\end{equation}
These notably remind us that the GR relation between the kinetics of the scalar and of the expansion is rather simple compared to here where,
 by addition, \begin{equation}\dot\phi_b^2 = \Big(2\varepsilon_1-w(1+\varepsilon_1-\sigma)\Big)\Mpl^2 H^2.  \end{equation}

The linear theory follows quickly using the \yl{EFT action} equations with the following identification of the $\alpha$ parameters to plug in previous eqn.~\eqref{eq:linEq}
\begin{equation} \Big(\aK, \aH, \aB, \aT, \aM \Big) = \Bigg( \frac{\dot\phi_b^2}{(1-w)\Mpl^2H^2},\, 0,\, \frac{w}{2(w-1)},\, -2 \aB,\, -\sigma \aT  \Bigg) \end{equation}
where the second comes from Horndeski membership of EGB gravity, the following two coming from the matching of $\psi^{\rm co}$ and $\chi^{\rm co}$, $\aT$ using the definition of $z$ with $f=1$ and $\aM$ from its definition.
These are consistent with the literature \cite{Gleyzes13, Gleyzes14} and yield
$$\frac{\abar}{(1+\aB)^2} = \frac{(1-w)}{(1-\frac{3}{2}w)^2}\Bigg( \frac{\dot\phi_b^2}{\Mpl^2 H^2}+\frac{3w^2}{2(1-w)}\Bigg) = \frac{(1-w)^2}{(1-\frac{3}{2}w)^2}\Bigg( \aK+\frac{3w^2}{2(1-w)^2}\Bigg),$$
leading us to write the Stochastic EGB equations of motion as
\begin{equation}
  \begin{aligned}
  G_{\mu\nu} - \frac{1}{\Mpl^2}\Big(T^{GB}_{\mu\nu} + T^{\phi}_{\mu\nu}\Big) & =  \frac{(1-w)^2}{2(1-\frac{3}{2}w)^2}\Bigg( \aK+\frac{3w^2}{2(1-w)^2}\Bigg)
                                   \tilde{\rm P}_{\mu\nu}\boldsymbol{{\cal S}}^{(0)} +\boldsymbol{{\cal S}}_{\mu\nu}^{(2)},\\
    \nabla_\mu\nabla^\mu\phi -\frac{1}{8}\lambda_{, \phi}(\phi){\cal G} - V_{, \phi}(\phi) & = -\sqrt{\aK}\Mpl\frac{(1-w)^{\frac{3}{2}}}{2(1-\frac{3}{2}w)^2}\Bigg(1+\frac{3w^2}{2(1-w)^2{\aK}}\Bigg)
                                   \boldsymbol{{\cal S}}^{(0)}, \\
  \end{aligned}
\end{equation}
where we have left implicit that any field $g_{\mu\nu}$ or $\phi$ on the lhs should be taken as their IR part, $g_{\mu\nu}^>$ or $\phi^>$. We will keep this implicit in the rest of the examples for readability.

Let us comment on the result compared to \cite{aldabergenov_towards_2025}. Overall, the structure of the equations is similar in terms of the coefficients but rather different in terms of the sources.
However, as discussed in Section~\ref{sec:cons}, the original constraints are still satisfied in our case, which is not the case in the mentioned work and could point out an inconsistency.
% This is a consequence of coarse-graining in a specific gauge, which we avoid by using gauge-invariant combinations of functionals of $W\R$ in the coarse-graining procedure. 
% Moreover, several authors including the cited work use the prescription $\Pi_k\mapsto W_k\Pi_k$ for the coarse-graining of momenta (notably for the field), which seems rather contradictory with derivations of stochastic inflation from path-integral methods,
% where integrating out the UV is done one the field, so that, equivalently, $\Pi_k\mapsto W_k\Pi_k+\dot{W}_k\phi_k$ instead. \yl{I thought we had established that this is not the problem?}

\subsection{Stochastic non-minimally coupled dynamics\label{sec:NMC}}
\yl{Following and building on historical approaches beyond GR}, we build the stochastic equations for non-minimally coupled theories, starting with subcases such as $f(R)$ theories.

\paragraph{$f(R)$ dynamics.}
We start with $f(R)$ Inflation. The action is
\begin{equation}
  {\rm S} =  \frac{\Mpl^2}{2}\int d^4x \sqrt{-g} f(R).
\end{equation}
By introducing an auxiliary field $\varphi$, we can rewrite the action as
\begin{equation}
  S =\frac{\Mpl^2}{2} \int d^4x \sqrt{-g} \left[  f'(\varphi)(R - \varphi) + f(\varphi) \right].
\end{equation}
On shell, one has $\varphi = R$ and thus the two actions are equivalent. Finally, by defining $\phi = f'(\varphi)$, we can rewrite the action as that of a dilaton
\begin{equation}
  S =\frac{\Mpl^2}{2}  \int d^4x \sqrt{-g} \left[ \phi R - U(\phi) \right],
\end{equation}
where the potential $U(\phi)$ is given by $ U(\phi) = f'(\varphi(\phi))\varphi(\phi)- f(\varphi(\phi)) $.
In the unitary gauge and since $M^2 = \Mpl^2 \phi$, we identify the match to the \yl{EFT action} parameters as\footnote{Integration by part is useful to find $\aB$.} %GHY term?
$$ \Big(\aK, \aH, \aB, \aT, \aM \Big) = \Bigg( 0,\,0,\,- \frac{1}{2}\frac{\dot\phi_b}{H\phi_b},\,0,\, \frac{\dot\phi_b}{H\phi_b} \Bigg), $$
noting the known absence of kinetics in the field. 
Using our main result, the stochastic equations are thus 
\begin{equation}
  \begin{aligned}
  \phi G_{\mu\nu}[g]  + \Big(g_{\mu\nu}\Box - \nabla_\mu\nabla_\nu\Big)\phi+\frac{1}{2}U(\phi)g_{\mu\nu} &= \frac{3\phi_b}{\Big(1-\frac{2H\phi_b}{\dot\phi_b} \Big)^2}\tilde{\rm P}_{\mu\nu}\boldsymbol{{\cal S}}^{(0)} + \phi_b \boldsymbol{{\cal S}}_{\mu\nu}^{(2)}\,,\\
  R[g] - U_{,\phi}(\phi) & = -\frac{12H\phi_b\dot\phi_b}{\Big(1-\frac{2H\phi_b}{\dot\phi_b} \Big)^2}\boldsymbol{{\cal S}}^{(0)}\,.
  \end{aligned}
\end{equation}
which shows the robustness of the method as it can be applied to a theory with no canonical kinetic term.
As we are about to see, this theory is a subcase of Brans-Dicke Inflation, which can be moved to Einstein frame and can thus be looked at via standard GR stochastic Inflation.

\paragraph{Damour-Esposito-Farèse dynamics.} Damour-Esposito-Farèse (DEF) gravity and thus inflation adds noncanonical kinetics to $f(R)$ inflation\footnote{Note that this action uses a dimensionless scalar field.}
\begin{equation}
  S = \int d^4x \sqrt{-g} \left[ \frac{\Mpl^2}{2} \Big(\phi R - \frac{\omega(\phi)}{2\phi}\nabla^\mu\phi\nabla_\mu\phi\Big) - V(\phi) \right],
\end{equation}
where $\omega$ is the generalised dimensionless Brans-Dicke parameter that controls the strength of the kinetic term, $0$ in the case of $f(R)$ gravity. When $\omega$ is constant, this is the Brans-Dicke theory.
The identification to the \yl{EFT action} parameters is
\begin{equation} \Big(\aK, \aH, \aB, \aT, \aM \Big) = \Bigg( {\omega(\phi_b)}\bigg(\frac{\dot\phi_b}{H\phi_b}\bigg)^2,\,0,\,- \frac{1}{2}\frac{\dot\phi_b}{H\phi_b},\,0,\, \frac{\dot\phi_b}{H\phi_b} \Bigg), \end{equation}
leading to the stochastic equations
\begin{equation}
  \begin{aligned}
  \phi G_{\mu\nu}  + \Big(g_{\mu\nu}\Box - \nabla_\mu\nabla_\nu\Big)\phi -\frac{\omega(\phi)}{2\phi} T^{\phi}_{\mu\nu}[\phi]& \\
   +\frac{1}{\Mpl^2}\bigg(1-\frac{\omega(\phi)}{\phi}\bigg)V(\phi)g_{\mu\nu} &= \phi_b\frac{3+2w_b}{\Big(1-\frac{2H \phi_b}{\dot \phi_b} \Big)^2}\tilde{\rm P}_{\mu\nu}\boldsymbol{{\cal S}}^{(0)}+\phi_b\boldsymbol{{\cal S}}_{\mu\nu}^{(2)},\\
\frac{\omega(\phi)}{\phi}\nabla^\mu\nabla^\mu\phi + \frac{1}{2}\left(\frac{\omega(\phi)}{\phi}\right)_{,\phi}\nabla^\mu\phi\nabla_\mu\phi + R - \frac{2}{\Mpl^2}V_{,\phi}(\phi)  &= -4H \phi_b\dot\phi_b\frac{3+2w_b}{\Big(1-\frac{2H \phi_b}{\dot \phi_b} \Big)^2}\boldsymbol{{\cal S}}^{(0)}.
  \end{aligned}
\end{equation}

\paragraph{Non-minimally coupled dynamics.}
The previous action is equivalent to the non-minimally coupled (NMC) one with coupling functional $F(\phi)>0$,
\begin{equation}
  S = \int d^4x \sqrt{-g} \left[ F(\phi) R - \frac{1}{2}\nabla^\mu\phi\nabla_\mu\phi - V(\phi) \right],
\end{equation}
which has a slighlty different identification to the \yl{EFT action} parameters as
\begin{equation} \Big(\aK, \aH, \aB, \aT, \aM \Big) = \Bigg( \frac{\dot\phi_b^2}{2FH^2},\,0,\,-\frac{1}{2} \frac{\dot F_b}{H F_b},\,0,\, \frac{\dot F_b}{H F_b} \Bigg), \end{equation}
and directly leading us to the stochastic equations
\begin{equation}
  \begin{aligned}
  F^> G_{\mu\nu}^> + \Big(g_{\mu\nu}^>\Box - \nabla_\mu\nabla_\nu\Big)F^> -\frac{1}{2} T^{\phi}_{\mu\nu}[\phi^>] & = \frac{6F_b+\frac{\dot\phi_b^2F_b^2}{\dot F_b^2}}{\Big(1-\frac{2H F_b}{\dot F_b} \Big)^2}\tilde{\rm P}_{\mu\nu}\boldsymbol{{\cal S}}^{(0)}+ F_b\boldsymbol{{\cal S}}_{\mu\nu}^{(2)},\\
 \nabla_\mu\nabla^\mu\phi^>+ F_{,\phi}(\phi^>)R^> - V_{,\phi}(\phi^>) & = -4H \dot\phi_b\frac{\Big(6F_b+\frac{\dot\phi_b^2F_b^2}{\dot F_b^2}\Big)}{\Big(1-\frac{2H F_b}{\dot F_b} \Big)^2}\boldsymbol{{\cal S}}^{(0)} .
  \end{aligned}
\end{equation}

% Dilaton only, ie noe extra field. Can be transformed to Einstein Frame and thus apply \cite{Launay24} with a simple potential.

% This includes both Brans-Dicke and f(R) theories. 

% Note, Salo's thesis gives the coefficient and its seems that G2 is w/phi-1 times X.

% That also shows that the alpha space is degenerate for single field (disformal transformation, which is a gauge transform, link some of the parameter space).

% Double check dilaton part with JGBellido94's paper.

% Can this be moved to the Einstein frame? Yes, using a conformal transformation.
 
\subsection{Stochastic Horndeski dynamics\label{sec:horn}}
The Horndeski theory is the most general nonlinear scalar-tensor theory with at most second-order derviatives equations of motion, which ensures the absence of Ostrogradsky instabilities. 
The associated action writes
\begin{equation}
{\rm S} =\int d^4x\sqrt{-g}\Big({\cal L}_2+{\cal L}_3+{\cal L}_4+{\cal L}_5\Big),
\label{eq:action_horndeski}
\end{equation}
where we parametrise the Lagrangian densities as
\begin{subequations}
  \begin{align}
      {\cal L}_2&=K(\phi, X),\\
{\cal L}_3&=-G_3(\phi, X)\Box\phi,\\
{\cal L}_4&=G_{4}(\phi, X){}^{(4)}R+G_{4,X}\left[
\left(\Box\phi\right)^2-\left(\nabla_\mu\nabla_\nu\phi\right)^2
\right],
\\
{\cal L}_5&=G_5(\phi, X) G_{\mu\nu}\nabla^\mu\nabla^\nu\phi
-\frac{1}{6}G_{5,X}\Bigl[
\left(\Box\phi\right)^3
-3\left(\Box\phi\right)\left(\nabla_\mu\nabla_\nu\phi\right)^2
+2\left(\nabla_\mu\nabla_\nu\phi\right)^3
\Bigr],
  \end{align}
\end{subequations}
with\footnote{Note that sometimes $X$ is defined without the $-1/2$ prefactor \cite{Gleyzes13}.\label{fn:Xnotations}}\edef\fnXnotationNum{\arabic{footnote}} $X:=-\nabla_\mu\phi \nabla^\mu\phi/2$ and $G_{i,Y}=\partial G_i/\partial Y$.
Since all previous examples are special cases of Horndeski theories, they can be matched to a choice of $K$, $G_3$, $G_4$ and $G_5$ functions.
This is given in Table \ref{tab:HornMatch}.

\setlength{\tabcolsep}{4pt}   
\renewcommand{\arraystretch}{1.7}
\begin{table}[h]
\begin{center}
\resizebox{\textwidth}{!}{%
\begin{tabular}{|l|c|c|c|}
\hline
Theory & $f$ & $c$ & $\Lambda$ \\
\hline
GR & $1$ & $\frac{\dot\phi_b^2}{2}$ & $V(\phi_b)$ \\
\hline
$f(R)$ & $\phi_b$ & $0$ & $\frac{\Mpl^2}{2}U(\phi_b)$ \\
\hline
DEF & $\phi_b$ & $\frac{\Mpl^2\,\omega(\phi_b)\,\dot\phi_b^2}{4\phi_b}$ & $V(\phi_b)$ \\
\hline
NMC & $\frac{2F(\phi_b)}{\Mpl^2}$ & $\frac{\dot\phi_b^2}{2}$ & $V(\phi_b)$ \\
\hline
EGB & $1$ & $\frac{\dot\phi_b^2}{2}$ & $V(\phi_b) + 3w\Mpl^2 H^2$ \\
\hline
Braiding & $1$ & $\frac{1}{2}\dot\phi_b^2\big(P_{,X}+(-\ddot\phi_b+3H\dot\phi_b)G_{,X}-2G_{,\phi}\big)$ & $-P+\frac{1}{2}\dot\phi_b^2\big(-P_{,X}+(\ddot\phi_b+3H\dot\phi_b)G_{,X}\big) $ \\
\hline
General Horndeski & $\frac{2}{\Mpl^2}G_4-\frac{\dot\phi_b^2}{\Mpl^2}G_{5,\phi}-\frac{\dot\phi_b^2\ddot\phi_b}{\Mpl^2}G_{5,X}$ & \multicolumn{2}{c|}{Appendix C of \cite{Gleyzes13}\footnotemark[\fnXnotationNum]} \\
\hline
\end{tabular}
}
\end{center}
\caption{Background EFT coefficient matching for each theory covered in Section~\ref{sec:EOMs}.
$f$ is the gravitational coupling prefactor ($M^2 = \Mpl^2 f/(1+\aT)$),
$c$ the kinetic coefficient, and $\Lambda$ the potential-like term in the background action~\eqref{eq:EFTaction}.}
\label{tab:bgMatch}
\end{table}

\setlength{\tabcolsep}{2pt}   
\renewcommand{\arraystretch}{1.7}
\begin{table}
\begin{center}
  % \resizebox{\textwidth}{!}{%
  %  \tiny
\begin{tabular}{|c|c|c|c|c|c|}
\hline
Action & $\aK$ & $\aB$ & $\aT$ & $\aM$ & $\aH$ \\
\hline
GR & $2\varepsilon_1$ & $0$ & $0$ & $0$ & $0$ \\
\hline
EGB & $\frac{\dot\phi_b^2}{(1-w)\Mpl^2H^2}$ & $\frac{w}{2(w-1)}$ & $\frac{w}{1-w}$ & $\frac{w\sigma}{w-1}$ & $0$ \\
\hline
$f(R)$ & $0$ & $-\frac{1}{2}\frac{\dot\phi_b}{H\phi_b}$ & $0$ & $\frac{\dot\phi_b}{H\phi_b}$ & $0$ \\
\hline
DEF & $\omega(\phi_b)\!\left(\frac{\dot\phi_b}{H\phi_b}\right)^2$ & $-\frac{1}{2}\frac{\dot\phi_b}{H\phi_b}$ & $0$ & $\frac{\dot\phi_b}{H\phi_b}$ & $0$ \\
\hline
NMC & $\frac{\dot\phi_b^2}{2F_bH^2}$ & $-\frac{\dot F_b}{2HF_b}$ & $0$ & $\frac{\dot F_b}{HF_b}$ & $0$ \\
\hline
Horndeski & \multicolumn{5}{c|}{see \cite{Gleyzes13,Gleyzes14}} \\
\hline
\end{tabular}
  % }
\end{center}
\caption{Summary of the $\alpha$ parameters of diverse theories, found through their quadratic action.}
\label{tab:alphaMatch}
\end{table}

\begin{table}
\begin{center}
  \resizebox{\textwidth}{!}{%
\begin{tabular}{|c|c|c|c|c|}
\hline
Action & $K$ & $G_3$ & $G_4$ & $G_5$ \\
\hline
GR & $X-V(\phi)$ & $0$ & $\frac{\Mpl^2}{2}$ & $0$ \\
\hline
EGB & $X-V+8X^2(3-\ln X)\partial_\phi^4\lambda$ & $\frac{\Mpl^2}{2}+4X(7-3\ln X)\partial_\phi^3\lambda$ & $4X(2-\ln X)\partial_\phi^2\lambda$ & $-4\ln|X|\partial_\phi\lambda$ \\
\hline
$f(R)$ & $-\frac{\Mpl^2}{2}U(\phi)$ & $0$ & $\frac{\Mpl^2}{2}\phi$ & $0$ \\
\hline
DEF & $\frac{\Mpl^2\,\omega(\phi)}{2\phi}X - V(\phi)$ & $0$ & $\frac{\Mpl^2}{2}\phi$ & $0$ \\
\hline
NMC & $X - V(\phi)$ & $0$ & $F(\phi)$ & $0$ \\
\hline
Braiding & $P(\phi, X)$ & $G$ & $\frac{\Mpl^2}{2}$ & $0$ \\
\hline
\end{tabular}
  }
\end{center}
\caption{Matching Horndeski functions to specific theories, see \cite{KYY11,Gleyzes13} for EGB.}
\label{tab:HornMatch}
\end{table}

There is also a correspondence between the $\alpha$ parameters and the $K$, $G_3$, $G_4$ and $G_5$ functions (using Appendix C of \cite{Gleyzes13} and Table 1 of \cite{Gleyzes14}),
which we could use to identify the source factors. 
Instead and for the sake of demonstration, we bypass the full identification of the $\alpha$ parameters using the derivation of the MS action found in \cite{KYY11}, giving us instead
\begin{subequations}
\begin{align}
  M^2(1+\aT)&=2\left[G_4
-X\left( \ddot\phi G_{5,X}+G_{5,\phi}\right)\right],
\\
M^2&=2\left[G_4-2 XG_{4,X}
-X\left(H\dot\phi G_{5,X} -G_{5,\phi}\right)\right],\\
z^2&=2a^3\bigg(\frac{\Sigma}{\Theta^2}M^4+3M^2\bigg),\\
c^2_s&= \frac{\frac{1}{a}d_t\Big(\frac{\Theta}{a}M^2\Big)-M^2(1+\aT)}{\frac{\Sigma}{\Theta^2}M^4+3M^2}
\label{eq:KobayashiM}
\end{align}
\end{subequations}
where $\Sigma$ and $\Theta$ are defined as
\begin{subequations}
\begin{align}
\Sigma&:=XK_X+2X^2K_{,XX} \notag\\
& +12H\dot\phi XG_{3,X}+6H\dot\phi X^2G_{3,XX} -2XG_{3,\phi}-2X^2G_{3,\phi X}\notag \\
& -6H^2G_4 +6\Bigl[H^2\left(7XG_{4,X}+16X^2G_{4,XX}+4X^3G_{4,XXX}\right) \notag\\
& -H\dot\phi\left(G_{4,\phi}+5XG_{4,\phi X}+2X^2G_{4,\phi XX}\right)
\Bigr]
\notag\\
& +30H^3\dot\phi XG_{5,X}+26H^3\dot\phi X^2G_{5,XX}
\notag\\&
+4H^3\dot\phi X^3G_{5,XXX}
-6H^2X\bigl(6G_{5,\phi}
+9XG_{5,\phi X}+2 X^2G_{5,\phi XX}\bigr),
\\
\Theta&:=-\dot\phi XG_{3,X}+
2HG_4-8HXG_{4,X}
-8HX^2G_{4,XX}+\dot\phi G_{4,\phi}+2X\dot\phi G_{4,\phi X}
\notag\\&
-H^2\dot\phi\left(5XG_{5,X}+2X^2G_{5,XX}\right)
+2HX\left(3G_{5,\phi}+2XG_{5,\phi X}\right).
\end{align}
\end{subequations}
In the past two blocks of equations, all quantities are implictly evaluated on the background trajectory, as usual within the stochastic ansatz-- if no heuristic for the stochastic backreaction is used.

Eqn.\eqref{eq:KobayashiM} is enough to deduce $\Src_{\cal R}^{\rm EFT}$ and $\Src_h^{\rm EFT}$ requiring $(z, c_s,c_T)$ as well as
\begin{equation}
\frac{\abar}{(1+\aB)^2}= \frac{z^2}{a^3M^2} = 2\bigg(3+\frac{\Sigma}{\Theta^2}M^2\bigg).
\end{equation}
One can check that we recover $2\varepsilon_1$ in the GR limit.

The Einstein and scalar-field equations then follow naturally using the EOMs from \cite{KYY11}
\begin{equation}
\begin{aligned}
\sum_{i=2}^5{\cal G}_{\mu\nu}^i &= \frac{z^2}{4a^3}\tilde{\rm P}_{\mu\nu}\boldsymbol{{\cal S}}^{(0)}+\frac{M^2}{2}\boldsymbol{{\cal S}}_{\mu\nu}^{(2)},\\
\quad \nabla^\mu\left(
\sum_{i=2}^5 J_\mu^i \right)-\sum_{i=2}^5 P^i_\phi&= -\frac{\dot\phi_bHz^2}{a^3}\boldsymbol{{\cal S}}^{(0)},
\end{aligned}
\end{equation}
where the explicit expressions of ${\cal G}_{\mu\nu}^i$, $J_\mu^i$ and $P_\phi^i$ are recalled in \yl{Appendix~\ref{app:horndeski}}.

\subsection{Stochastic Braiding dynamics} With the general case of stochastic Horndeski theories treated, it is easier to access Braiding dynamics.
Kinetic Braiding is the most general Hornesdki with nonzero $K=P$ and $G_3=G$ \footnote{Note that we do not use a minus sign.} but GR's $G_4=\Mpl^2/2$ and $G_5=0$, or, in the $\alpha$-basis language, nontrivial $\aK$ and $\aB$ but null $\aT$ and $\aM$. 

This leads to the stochastic equations of motion
\begin{equation}
  \begin{aligned}
   G_{\mu\nu} -\frac{1}{\Mpl^2}T_{\mu\nu}^{\rm Braiding}
      & = C_b\tilde{\rm P}_{\mu\nu}\boldsymbol{{\cal S}}^{(0)}+\boldsymbol{{\cal S}}_{\mu\nu}^{(2)},\\
    \nabla^{\mu}\Bigg(\Big((G\Box\phi)_X-P_X+2G_\phi\Big)\nabla_{\mu}\phi +G_X\nabla_{\mu}X \Bigg)-\nabla^{\mu}G_\phi\nabla_{\mu}\phi-P_\phi&= -2\dot\phi_b HC_b \Mpl^2\boldsymbol{{\cal S}}^{(0)},
    \end{aligned}
\end{equation}
where the stress-energy tensor is given by
\begin{equation}
    T_{\mu\nu}^{\rm Braiding} = P_{X} \nabla_{\mu}\phi\nabla_{\nu}\phi 
    + \Big(P +\nabla_{\lambda}G\nabla^{\lambda}\phi\Big) g_{\mu\nu} -G_{,X}\Box\phi \nabla_{\mu}\phi\nabla_{\nu}\phi 
     -2\nabla_{\mu}G\nabla_{\nu}\phi,
\end{equation}
and the background front factor
\begin{equation} C_b \equiv 3+\frac{XP_X+2X^2P_{XX}+12H\dot\phi_b XG_{,X}+6H\dot\phi_b X^2G_{,XX}-2XG_{,\phi}-2X^2G_{,\phi X}-3H^2\Mpl^2}{H-\dot{\phi}_bXG_X/\Mpl^2}.\end{equation}

As identified in the literature \cite{deffayet_imperfect_2010}, the scalar field equation receives a contribution from second order derivatives of the metric via the term $\nabla^{\mu}\nabla_{\mu}X$
 and can thus be simplified using the Einstein equation, here including the stochastic sources. This is a hallmark of Horndeski theories as explained in \cite{KYY11}, as part of the features ensuring that the EOMs stay at most second order in derivatives.
 
\section{Stochastic ansatz at the edge? \label{sec:discussion}}

In the following, we discuss some further insights and extensions before concluding.

\subsection{A word on the EFT of Inflation}

% The EFT of dark energy, as we have seen \yl{through the lens of our notations}, completely includes the inflaton, and thus the EFT of inflation.
Recently \cite{Choudhury25}, stochastic Langevin equations were derived in a model-independent single-field EFToI framework, using it as a model-agnostic way to look at primordial black hole production
\begin{equation}
  \begin{aligned}
    \frac{\d\hat{\zeta}}{\d\cN} & = \hat{\Pi}_{\zeta} + \hat{\xi}_{\zeta}(\cN), \\
    \frac{\d\hat{\Pi}_{\zeta}}{\d\cN} & = -(3-\epsilon)
    \left(1 - \frac{2\!\left(s - \frac{\eta}{2}\right)}{3-\epsilon}\right)\hat{\Pi}_{\zeta}
  + \hat{\xi}_{\pi\zeta}(\cN), \label{eq:Choudhury}
  \end{aligned}
\end{equation}
where $s = \frac{\d\ln c_s}{\d\cN}$, $\;\epsilon = -\frac{\d\ln H}{\d\cN}$,
$\;\eta = \epsilon - \frac{1}{2}\frac{d\ln\epsilon}{d\cN}$, and $\xi_{\zeta}$ and $\xi_{\pi\zeta}$ are the Gaussian noises from the UV curvature perturbation and its conjugate momentum, respectively.
Compared to our general treatment, this equation focuses on
1) the long wavelength limit
2) the terms of the EFT giving $c_s\neq 1$ theories (so that the quadratic action has $f=1$, $\aB=\aM=\aT=\aH=0$)
3) the linear regime of $\zeta$ on those scales (equating to using eqn.\eqref{eq:MS} in the limit of 1) and 2)).

We do not make any claims on the validity of those approximations.
Given the uncertainty on the matter from the cited work, we bring to the attention of the reader that,
indeed some non-Gaussianity would arise in the tail of the distribution even within this linear framework, as it emerges from the first-passage time structure of the Fokker-Planck problem, or the $\Delta {\cal N}$ formalism (stochastic or not): the distribution of exit efold times $\mathcal{N}$ is non-Gaussian even when driven by Gaussian noise and a linear drift, because the question of when a trajectory first reaches the threshold $\zeta_{\rm th}$ is a nonlinear functional of the entire stochastic path.
However, one might want to justify the use of the linear regime, as both the Stueckelberg trick and the conversion from $\pi$ to $\zeta$ are nonlinear operations.
In fact, even if quantum diffusion, i.e. stochastic effects, were to dominate drift and other nonlinear parts of the evolution (such as those enforced by constraints), it is not clear that the linear LHS of eqn.~\eqref{eq:Choudhury} would indeed stay so.
Ultra slow roll dynamics might help \yl{simplify} dynamics drastically but that is \textit{a priori} not trivial and needs justification.

Besides the quoted study, it is possible to find a fully nonlinear and nonperturbative equation for $\pi$ in a model-agnostic way.
This follows from evaluating the EFT in the decoupling limit \cite{Creminelli25} to get the $\pi$ EOM
\begin{equation}
  \ddot{\pi} = -\frac{\ddot{H}}{2\dot{H}}\Bigg|_{t+\pi}\Bigg(\dot{\pi}^2-\frac{(\partial_i)^2\pi}{a^2}\Bigg).
\end{equation}
Thus if we were to formulate stochastic inflation for $\pi$ in this limit, we would linearise this equation and coarse-grain it.
Interpreting this result could help \yl{understand} the link between stochastic inflation and nonperturbative methods using the saddlepoint of the decoupled wavefunctional \cite{creminelli_dissipative_2023}, as well as potentially providing an actual stochastic EFT framework--- as opposed to using an EFT as a wide class of theories to apply the stochastic formalism to, as we do in this work.
This is beyond the scope and interests of this work.

\subsection{Numerical evolution}
Of course, one is free to investigate the long wavelength approximation of the equations provided in this work, or the decoupling limit.
Given the recently found caveats of the separate universe approximation regarding e.g. the account of gauge and gradients effects and the numerous heuristics it takes to partially fix them \cite{artigas_hamiltonian_2022, jackson23, briaud_stochastic_2025},
this research advocates for a fully numerical approach once the most 
general equations have been derived, following insights developed 
in our previous work on general relativity and numerical relativity 
(NR)~\cite{Launay24, Launay2025}.

Moreover, deriving the full Horndeski stochastic equations comes at an opportune 
time: the first fully nonlinear NR simulations of inhomogeneous 
inflation beyond GR have recently been carried out using 
$\mathsf{GRFolres}$~\cite{areste_salo_puncture_2023, brady_inflaton_2025}. These simulations rest on 
a series of works establishing the well-posedness of classes of Horndeski theories 
in suitable gauges and variables~\cite{kovacs_well-posed_2020,salo_well-posedness_2022,salo_strong_2024}.\footnote{Well-posedness seems to be currently guaranteed within the weak-coupling regime
for scalar-tensor theories, though recent work has extended this to
all polynomial vacuum EFTs beyond weak coupling~\cite{figueras_well-posed_2024} \yl{(preprint)}.}
In the future, and given the implementation of \cite{brady_inflaton_2025} within the same software as our previous work in NR \cite{Launay2025}, and given the simplicity of the source conversion to the numerical variables \cite{Launay24} (i.e. going from Einstein's equations to ADM/\yl{Baumgarte-Shapiro-Shibata-Nakamura} ones), stochastic inflation beyond GR is ready to provide its first assumption-free estimates in the classical regime.
Signatures from higher derivatives \yl{are} rather unclear and limited in expectations, thus reaffirming the motivations of such numerical and general studies.
However, in the proxy built in \cite{Choudhury25} following established derivations \cite{vennin_correlation_2015}, one sees that a nontrivial speed of sound \yl{alone} should affect the drift in $\zeta$ and therefore the PDF of $\cN$.
As a consequence, the PBH abundance computed from the tail of the PDF will vary greatly, as well as the associated correlators. \yl{Conversely}, measuring PBHs or not, could help \yl{constrain} corrections to GR in the early universe.

\subsection{Beyond the \yl{EFT action}}
Going beyond theories described by the \yl{EFT action} seems rather ambitious, at first sight. 
In fact, some of the theories already included in this work showcase rather complicated dynamics, whether we aim at finding the stochastic equations or solving them numerically.
These include beyond Horndeski theories ($\aH\neq 0$), which have higher-order time derivatives in the equation of motion and which we have not looked at in our examples of Section~\ref{sec:EOMs}.
Thanks to recent studies, higher-order time derivative theories and the presence of ghosts have been disentangled dramatically to go beyond second-order relativity in time \cite{gleyzes_healthy_2015}. 
This made \yl{it possible to include} $\aH\neq 0$ in the present formalism, but also to look at further classes, including (Unified) Degenerate Higher-Order Scalar-Tensor --- so-called GPLV and (U)DHOST --- theories \cite{LangloisNoui16,DeFelice18} and recent extensions \cite{Gorji26, Michiwaki26}.
One can imagine that the coarse-graining of such theories would be more \yl{involved} (more time derivatives to handle) but not impossible with our procedure --- in principle.

The point being raised here is indirectly the question of whether or not one needs to know the $\alpha$ coefficients.
Clearly we have shown in the Horndeski case in Section~\ref{sec:horn} that all one needs to find the sources is to know the prefactors in front of the scalar and gravitonic Mukhanov-Sasaki actions.
Thus, beyond the validity of the stochastic limit, one could imagine that this nonlinear to linear correspondence of the coefficients is all one needs to find the stochastic equations, and so should not be limited to dynamically-friendly dynamics.

\subsection{Beyond the Inflaton: multifields and all \label{sec:beyondinflaton}}
In general, scalar tensor theories used for inflation do not have to use that particular scalar as the inflaton,
and more generally, one should be considering the possibility of a much richer phenomenology of all spins surrounding, in the case of inflation, an overall vacuum dark-energy-like period of time.
Importantly, a frame might have a multifield action with \yl{field-space} curvature and this needs to be accessible.

Starting with a spin-0 landscape, our formalism --- and subsequent NR simulations --- can be generalised to multifields.
On the one hand, its EFT has already been written \cite{senatore_effective_2012}.
On the other hand, Stochastic Inflation already has a consequent literature, though in the long wavelength regime \cite{BattefeldVanTent06, Rigopoulos_multi06, Takahashi26}.
Let us sketch a procedure for the coarse-graining of multifields GR, with field-space metric ${\rm G}_{IJ}(\phi^K)$, potential $V(\phi^I)$, and action\footnote{See \cite{Sasaki98, nibbelink_scalar_2002} for further introduction to the formalism.}
\begin{equation}
    S = \frac{\Mpl^2}{2}\int d^4x \sqrt{-g} R - \int d^4x \sqrt{-g}\Bigg( \frac{1}{2} {\rm G}_{IJ}\nabla_\mu\phi^I\nabla^\mu\phi^J+V(\phi)\Bigg),
\end{equation}
yielding the nonlinear EOMs
\begin{subequations}
  \begin{align}
    \Box {\phi}^I + \Gamma^{I}_{JK}\nabla_{\mu}\phi^J\nabla^\mu\phi^K + {\rm G}^{IJ}\partial_J V & = 0, \\
    G_{\mu\nu} - \frac{1}{\Mpl^2}{\rm G}_{IJ}\nabla_\mu\phi^I\nabla_\nu\phi^J + \frac{1}{\Mpl^2} g_{\mu\nu}\left(\frac{1}{2}{\rm G}_{IJ}\nabla^\alpha\phi^I\nabla_\alpha\phi^J + V\right) &= 0,
  \end{align}\label{eq:multifields}
\end{subequations}
where $\Gamma^{A}_{BC}$ are the Christoffel symbols of the field-space metric ${\rm G}$. Compared to GR, all components are simply modified through the effective stress-energy tensor. Our procedure would then, as usual, \yl{proceed as follows:}
\begin{enumerate}
\item \underline{Identify linear gauge-invariants} such as
            \begin{subequations}
            \begin{align}
              Q^I \equiv \delta \phi^I +\frac{\dot{\phi}_b^I}{H}\Phi
            \end{align}
            \end{subequations}
            and pick one to coarse-grain (here $Q^I$, following a generalisation of our approach with ${\cal R}$ but that is not compulsory).
            $Q^I$ has a known matrix-mass Mukhanov-Sasaki equation of the form
            \begin{equation}
              \mathcal{D}_t^2 Q^I + 3H\,\mathcal{D}_t Q^I - \frac{\nabla^2}{a^2}Q^I + \mathcal{M}^I{}_J\, Q^J = 0, \label{eq:multiMS}
            \end{equation}
            where $\mathcal{D}_t (\cdot)^I \equiv \partial_t {(\cdot)}^I + \Gamma^I_{JK}\dot\phi^J_b (\cdot)^K$ is the field-space covariant time derivative.
\item \underline{Express all perturbations of one gauge in terms of one of them} ($Q^I$).
      Using the constraints with their new stress-energy contributions
            \begin{subequations}
              \begin{align}
                \delta\rho &= {\rm G}_{IJ}{\dot{\phi}^I_b} {\delta\dot{\phi}^J}
                + \partial_I V\,\delta\phi^I
                - {\rm G}_{IJ}{\dot{\phi}^I_b \dot{\phi}^J_b}\Psi, \\
                \delta\mathcal{J}_i &= -{\rm G}_{IJ}{\dot{\phi}^I_b}\,\partial_i\delta\phi^J.
              \end{align}
            \end{subequations}
            These terms are particularly special because they, apart from the $\partial_I V$ contribution, constitute the adiabatic contributions (meaning along tangent to the background field-space trajectory).
            The multifield comoving gauge thus generalises naturally to nulling the adiabatic sources ${\rm G}_{IJ}{\dot{\phi}^I_b}\,\delta\phi^J=0$ and the usual $E=0$, as done in Section~\ref{sec:slicingsOfR} to get all perturbations in terms of the gauge invariant quantity.

            However the constraints cannot be solved on their own in this gauge as opposed to the single-field case (entropic contributions of the fields make the \yl{Hamiltonian} constraints nontrivial),
            which is why the flat gauge ($\Phi=E=0$, $\delta\phi^I = Q^I$) can be used to solve the constraints and extract $\Psi$ from the momentum constraint, and $B$ from the \yl{Hamiltonian} constraint using that $\Psi$. We find
            \begin{subequations}
                \begin{align}
                  \delta \phi^{I,{\rm flat}} & = Q^I,\\
                  E^{\rm flat} & = 0,\\
                  \Psi^{\rm flat} & = \frac{1}{\Mpl}\sqrt{\frac{\epsilon}{2}}Q, \\
                  B^{\rm flat} & = \frac{1}{2a^2k^2\Mpl^2}\Big(\sqrt{2\epsilon}H\Mpl\big(\dot{Q}\big)_\parallel+\partial_I V_b Q^I + \sqrt{2\epsilon}(3-\epsilon)\frac{\Mpl}{H}Q_\parallel\Big)\\
                              & = -a^{-2} \chi^{\rm flat}
                \end{align}
            \end{subequations}
          where we defined the multifield slow-roll $\epsilon = G_{IJb}\dot{\phi}^I_b\dot{\phi}^J_b /(2 H^2 \Mpl^2)$ and the field-space adiabatic component projection $X_\parallel \equiv ({\rm G}_{IJ}\dot\phi_b^IX^J)/\sqrt{{\rm G}_{IJ}\dot\phi_b^I\dot\phi_b^J}$.
          This also gives us access to other generalised gauge invariant combinations
          \begin{subequations}
            \begin{align}
            \Phi_B     &\equiv \Phi + H\chi = H\chi^{\rm flat}[\boldsymbol{Q}]\,, \\
            \Psi_B     &\equiv \Psi - \dot\chi = \Psi^{\rm flat}[\boldsymbol{Q}] - \dot\chi^{\rm flat}[\boldsymbol{Q}]\,, \\
            Q^{I,{\rm gi}} & \equiv \delta\phi^I -  \sqrt{2\epsilon}H\Mpl\chi =Q^I -\sqrt{2\epsilon}H\Mpl\chi^{\rm flat}[\boldsymbol{Q}]\,, \\
            \zeta^{\rm gi}  & \equiv -\Phi + \frac{1}{3}\frac{\delta\rho}{\rho+p} =  \frac{\delta\rho^{\rm flat}[\boldsymbol{Q}]}{6\epsilon H^2\Mpl^2}\,,
          \end{align}
          \end{subequations}
          from which we can express all perturbations in terms of $Q^I$ and another field, or even just $Q^I$ in a specific gauge, see Section~\ref{sec:slicingsOfR}.
\item \underline{Obtain the sources to the nonlinear equations}, eqn.~\eqref{eq:multifields}, by linearising them and applying $Q^I\longrightarrow Q^{J>}\equiv W^I_J Q^J$ in the previously found decompositions,
where $W$ is a windowing matrix allowing to coarse-grain depending on the crossing of each field.
Given our previous findings, sources should all be found proportional to that of eqn.~\eqref{eq:multiMS}.
The final equations, their extension beyond GR following Section~\ref{sec:procedure}, and their application \yl{are} kept for future work, e.g. using numerical relativity\footnote{In fact, there does not seem to be any NR code including field space curvature, only flat field-space approximations \cite{Joana22multi}.}.
\end{enumerate} 

% \yl{adiabaticity is broken for multiple fields and so can maybe have secular growth? as opposed to inflatonic inhomogeneities. cite paper? zaldagarria?}

Beyond spin-0, we could perfectly imagine a similar construction for multiple spin-2 fields \yl{as} we did previously or less commonly multiple spin-1 fields\footnote{One could even imagine building a \yl{multispin} field-space as well --- if that is possible and motivated.},
as long as we understand sufficiently the associated linear theory, their 2-point function, and their regime of classicality or separation of scales.

\subsection{Beyond Closed EFTs}
As a final direction for extension, it has become clear in the past few years \yl{\cite{Salcedo24, salcedo_open_2024-1, Dufner1, salcedo_phenomenology_2026, Salcedo26sdn}} that an open system EFT description in cosmology would be more appropriate if we can only measure some parts of a large system of scales and fields.
The link with stochastic inflation seems rather obvious at first sight, as one focuses on an open system of wavenumbers in the latter.
The so-called \textit{OpenEFT} framework is seemingly more general as it traces out higher-energy but more generally hidden degrees of freedom and is formulated within a fully quantum description, thus containing stochastic inflation as a limit.
Very recently, analytical links have been established between the two approaches \cite{Li26, Cespedes26, Li26b}, although limited to the decoupling limit, similarly to the EFToI framework.
We hope that this could be improved in the next few years before extending our framework. %\yl{Also scalars other than the inflaton trigger secular growth. different breaking of the perturbativity.}

\subsection{Beyond inflation \label{sec:beyondinflation}}

\yl{
Although focused on the coarse-graining of scalar-tensor theories of the inflaton, 
the purpose of this work is to demonstrate a generic framework for the coarse-graining of fluctuations in a gauge-consistent way, first proposed in \cite{Launay24}.
There are quite a few ways to go beyond inflation from the perspective of stochastic or coarse-grained cosmology, which we try to sketch here with no expertise beyond suggestion.

Firstly, one could simply follow the arrow of time starting from any result obtained in an inflationary calculation.
Our methods, if implemented similarly to the GR case \cite{Launay26}, enable the creation of nonlinear initial conditions for the post-inflationary universe, which can then be evolved using numerical relativity or other methods in the right regimes.
This includes initial conditions for large-scale structure and $N$-body or GR simulations (e.g.\ $\mathsf{gevolution}$~\cite{1509.01699}, $\mathsf{(G)RAMSES}$~\cite{astro-ph/0111367, 2010.08257}), reheating and preheating lattice codes (e.g.\ $\mathsf{CosmoLattice}$~\cite{2006.15122}), and numerical-relativity collapse codes used for primordial black hole formation (e.g. $\mathsf{GRChombo}$~\cite{2507.19166}).
Bridging stochastic inflation effects and late-time cosmology codes in this way will be a major step in augmenting the testability of inflationary models.

Secondly, the formalism developed in this work is specific neither to inflation nor to
a fundamental field. All the construction needs is a perturbative sector and
a classical nonlinear one, in the UV or the IR, not necessarily reflecting a
quantum-to-classical transition.

In the case of dark energy in scalar-tensor theories, the EFT fixes the sources once a background and the
$\alpha$-functions are chosen, and inflation enters only through that choice with initial conditions.
In particular, the benefit of using the notations of the EFToDE is that it allows us to treat post-inflationary DE fluctuations with the same formalism.
If we keep the Hubble radius as the coarse-graining scale, it is increasing (as opposed to inflation) and separates two mostly linear systems, with no particularly interesting dynamics.
Our stochastic DE scalar-tensor equations would thus be beneficial for the study of the large-scale structure if we set the nonlinearity scale $k_{\rm NL}(t)$ to be that of the window for the study of small scales.
We know, however, that the effect of DE on the growth of structure is mostly linear and that the nonlinearities are mostly due to the matter fluid~\cite{astro-ph/9906174}, since a bigger impact from dark energy fluctuations at the nonlinearity scale would require DE clustering to be significant, found in specific models.\footnote{One way to achieve that is to have a low sound speed~\cite{1101.1026}, which is the case for kinetic braiding~\cite{deffayet_imperfect_2010,1103.5360} or coupled models~\cite{astro-ph/9908023}, as long as the coupling is not screened in dense regions~\cite{astro-ph/0309300}.}
Regarding simulations, using the stochastic EOMs instead of simple stochastic initial conditions is only relevant if entering fluctuations are only accessible dynamically or emerging in time.
During inflation, this is the case because we can only use a classical treatment for modes crossing the dynamical window. For late-time DE, we expect that this could depend on the model.

This work is not restricted to DE fluctuations. In the coarse-graining of multifields in GR carried out in Section~\ref{sec:beyondinflaton}, we have clearly shown that fields (whether or not they include the inflaton) can be coarse-grained and that they can source nonlinear equations of spacetime.
In using different gauge-invariant combinations in the coarse-graining, we could similarly coarse-grain non-scalar fluids with the same procedure, using a windowing on hypersurfaces of constant energy density (as opposed to constant field value) to define the coarse-graining scale.
This would amount to using $ \zeta^{\rm gi}_I  \equiv -\Phi + \frac{1}{3}\frac{\delta\rho_I}{\rho_I+p_I} $ instead of $\R$ (or $Q^I$) and the hydrodynamical equations of the fluid instead of the scalar field equation. This could help refine the construction of the EFT of large-scale structures~\cite{Baumann_2012, carrasco_effective_2012}, and more generally,
the study of the FLRW metric in full nonlinear GR with corrections arising from small scales, and vice versa.
In particular, the backreaction of small-scale inhomogeneities on larger scales, whether tensorial or matter \cite{Isaacson1968a,Isaacson1968b,1011.4920}, could now be studied in a more systematic way with stochastic methods accounting for subscale physics.

The unbroken assumption left in this work is that of an FLRW background for the fluctuations.
If we were to extend the formalism to a more general background, such as compact objects including black holes, we would need to find a new set of gauge-invariant quantities to coarse-grain, which also seems reasonable to find.
Previous work on more generic stochastic gravity includes \cite{verdaguer_stochastic_2007} and could help us in this direction.

}

\subsection{Conclusion}

In this work, we studied the coarse-graining of perturbations in scalar-tensor theories of dark energy and beyond, including multifields.
Provided a well-defined classical set of scales and a perturbative one, we showed that the stochastic sources for the curvature perturbation and the geometric quantities can be expressed in terms of the Mukhanov-Sasaki source, with modifications arising from the specific features of the theory under consideration.
This result provides a unified framework for studying stochastic inflation in a broad class of theories and paves the way for future investigations into the phenomenological implications of these models. 

With this method and as we proved in this piece of work, there is fundamentally no obstacle to generalise the procedure to any theory \yl{with gauge degrees of freedom} as long as they admit a classical set of scales and a perturbative one, whether \yl{the latter} is classical or quantum.
%The true challenges are rather in subtleties of the assumptions, which we think pose the true challenge to the community, such as the stochastic backreaction and the systematic determination of the semiclassical portions of spacetime.

\acknowledgments
I thank Thomas Colas, Panagiotis Giannadakis, Enrica Lausdei, and Gerasimos Rigopoulos for their comments on the draft, as well as my supervisor Paul Shellard for his support.
\yl{I also thank the referees for their valuable and rigourous feedback.}
% I would also like to thank all individuals who helped or are helping me advertising this work in my travels this year:
% the cosmology group in Queen Mary University London, the 6th EPS conference at the University of Stavanger, 
I acknowledge funding from the Kavli Institute for Cosmology Cambridge and the STFC DiS CDT program.

I thank the institutions who welcomed me to present this work, during a tour of Europe by train funded by a travel grant from Wolfson College, University of Cambridge, and in most cases by additional funding from the hosts: the Institute of Cosmology and Gravitation, Université catholique de Louvain, the CERN Theory Department, the University of Pisa and INFN, ICTP Trieste, FZU, the Institute of Theoretical Physics of the Jagiellonian University, the Potsdam Albert Einstein Institute, and GRAPPA at the University of Amsterdam.
For their invitation, interest, and related discussions, I thank my hosts: David Wands, Eemeli Tömberg, Christophe Ringeval, Matteo Braglia, Giovanni Marozzi, Amelia Drew, Matthew Elley, George Zahariade, Guillermo Lara, and Angelo Caravano, as well as others for discussions: Luiz Felipe Demetrio, Parth Bhargava, Baptiste Blachier, Chiara Animali, Pierre Bechaz, Enrico Barausse, Mehrdad Mirbabayi, Panagiotis Giannadakis, Will Barker, Nicolas Loayza, Dražen Glavan, Daniel del-Corral, Héloïse Delaporte, Jan-Pieter van der Schaar, Ben Freivogel and Alessandro Fumagalli.

\newpage

\appendix
\section{Coefficients of the Horndeski EOMs}\label{app:horndeski}
The terms below were calculated in \cite{KYY11, gao_inflation_2011}

\begin{eqnarray}
{\cal G}_{\mu \nu}^2&=&-\frac{1}{2}K_X\nabla_\mu\phi\nabla_\nu\phi-\frac{1}{2}Kg_{\mu\nu},
\\
{\cal G}_{\mu \nu}^3&=&\frac{1}{2}G_{3,X}\Box\phi\nabla_\mu\phi\nabla_\nu\phi
+\nabla_{(\mu}G_3\nabla_{\nu)}\phi-\frac{1}{2}g_{\mu\nu}\nabla_\lambda G_3\nabla^\lambda\phi ,
\\
{\cal G}_{\mu \nu}^4&=&
G_4G_{\mu\nu}-\frac{1}{2}G_{4,X}R\nabla_\mu\phi\nabla_\nu\phi
-\frac{1}{2}G_{4,XX}\left[(\Box\phi)^2-(\nabla_\alpha\nabla_\beta\phi)^2\right]\nabla_\mu
\phi\nabla_\nu\phi
\cr&&
-G_{4,X}\Box\phi\nabla_\mu\nabla_\nu\phi
+G_{4,X}\nabla_\lambda\nabla_\mu\phi\nabla^\lambda\nabla_\nu\phi
+2\nabla_\lambda G_{4,X}\nabla^\lambda\nabla_{(\mu}\phi\nabla_{\nu)}\phi
\cr&&
-\nabla_\lambda G_{4,X}\nabla^\lambda\phi\nabla_\mu\nabla_\nu\phi
+g_{\mu\nu}\left(G_{4,\phi}\Box\phi-2XG_{4,\phi\phi}\right)
\cr&&
+g_{\mu\nu}\biggl\{-2G_{4,\phi X}\nabla_\alpha\nabla_\beta\phi\nabla^\alpha\phi\nabla^\beta\phi
+G_{4,XX}\nabla_\alpha\nabla_\lambda\phi\nabla_\beta\nabla^\lambda\phi\nabla^\alpha\phi\nabla^\beta\phi
\cr&&
+\frac{1}{2}G_{4,X}\left[(\Box\phi)^2-(\nabla_\alpha\nabla_\beta\phi)^2\right]\biggr\}
+2\Bigl[G_{4,X}R_{\lambda(\mu}\nabla_{\nu)}\phi\nabla^\lambda\phi
\cr&&
-\nabla_{(\mu}G_{4,X}\nabla_{\nu)}\phi\Box\phi\Bigr]
-g_{\mu\nu}\left[
G_{4,X}R^{\alpha\beta}\nabla_\alpha\phi\nabla_\beta\phi-\nabla_\lambda G_{4,X}\nabla^\lambda\phi
\Box\phi\right]
\cr&& +G_{4,X}R_{\mu\alpha\nu\beta}\nabla^\alpha\phi\nabla^\beta\phi
-G_{4,\phi}\nabla_\mu\nabla_\nu\phi-G_{4,\phi\phi}\nabla_\mu\phi\nabla_\nu\phi
\cr&&
+2G_{4,\phi X}\nabla^\lambda\phi\nabla_\lambda
\nabla_{(\mu}\phi\nabla_{\nu)}\phi
-G_{4,XX}\nabla^\alpha\phi\nabla_\alpha\nabla_\mu\phi
\nabla^\beta\phi\nabla_\beta\nabla_\nu\phi,
\\
{\cal G}_{\mu \nu}^5&=&
G_{5,X}R_{\alpha\beta}\nabla^\alpha\phi\nabla^\beta\nabla_{(\mu}\phi
\nabla_{\nu)}\phi
-G_{5,X}R_{\alpha(\mu}\nabla_{\nu)}\phi \nabla^\alpha\phi\Box\phi
\cr&&
-\frac{1}{2}G_{5,X}R_{\alpha\beta}\nabla^\alpha\phi\nabla^\beta\phi\nabla_\mu\nabla_\nu\phi
-\frac{1}{2}G_{5,X}R_{\mu\alpha\nu\beta}\nabla^\alpha\phi\nabla^\beta\phi\Box\phi
\cr&&
+G_{5,X}R_{\alpha\lambda\beta(\mu}\nabla_{\nu)}\phi\nabla^\lambda\phi
\nabla^\alpha\nabla^\beta\phi
+G_{5,X}R_{\alpha\lambda\beta(\mu}\nabla_{\nu)}\nabla^\lambda\phi\nabla^\alpha\phi\nabla^\beta\phi
\cr&&
-\frac{1}{2}\nabla_{(\mu}\left[G_{5,X}\nabla^\alpha\phi\right]\nabla_\alpha\nabla_{\nu)}\phi\Box\phi
+\frac{1}{2}\nabla_{(\mu}\left[G_{5,\phi}\nabla_{\nu)}\phi\right]\Box\phi
\cr&&
-\nabla_\lambda\left[G_{5,\phi}\nabla_{(\mu}\phi\right]\nabla_{\nu)}\nabla^\lambda\phi
\cr&&
+\frac{1}{2}\left[
\nabla_\lambda \left(G_{5,\phi}\nabla^\lambda\phi\right)-\nabla_\alpha\left(G_{5,X}\nabla_\beta\phi\right)
\nabla^\alpha\nabla^\beta\phi
\right]\nabla_\mu\nabla_\nu\phi
\cr&&
+\nabla^{\alpha}G_5\nabla^{\beta}\phi R_{\alpha(\mu\nu)\beta}
-\nabla_{(\mu}G_5G_{\nu)\lambda}\nabla^\lambda\phi
\cr&&
+\frac{1}{2}\nabla_{(\mu}G_{5,X}\nabla_{\nu)}\phi\left[
(\Box\phi)^2-(\nabla_\alpha\nabla_\beta\phi)^2\right]
\cr&&
-\nabla^\lambda G_5R_{\lambda(\mu}\nabla_{\nu)}\phi
+\nabla_\alpha\left[G_{5,X}\nabla_\beta\phi\right]\nabla^\alpha\nabla_{(\mu}\phi
\nabla^\beta\nabla_{\nu)}\phi
\cr&&
-\nabla_\beta G_{5,X}\left[
\Box\phi\nabla^\beta\nabla_{(\mu}\phi-\nabla^\alpha\nabla^\beta\phi
\nabla_\alpha\nabla_{(\mu}\phi\right]\nabla_{\nu)}\phi
\cr&&
+\frac{1}{2}\nabla^\alpha\phi\nabla_\alpha G_{5,X}\left[
\Box\phi\nabla_\mu\nabla_\nu\phi -\nabla_\beta\nabla_\mu\phi\nabla^\beta\nabla_\nu\phi\right]
\cr&&
-\frac{1}{2} G_{5,X} G_{\alpha\beta}\nabla^\alpha\nabla^\beta\phi\nabla_\mu\phi\nabla_\nu\phi
-\frac{1}{2}G_{5,X}\Box\phi\nabla_\alpha\nabla_\mu\phi\nabla^\alpha\nabla_\nu\phi
\cr&&
+\frac{1}{2}G_{5,X}(\Box\phi)^2\nabla_\mu\nabla_\nu\phi
+\frac{1}{12}G_{5,XX}\bigl[(\Box\phi)^3-3\Box\phi(\nabla_\alpha\nabla_\beta\phi)^2
\cr&&
+2(\nabla_\alpha\nabla_\beta\phi)^3\bigr]\nabla_\mu\phi\nabla_\nu\phi
+\frac{1}{2}\nabla_\lambda G_5 G_{\mu\nu}\nabla^\lambda\phi
\cr&&
+g_{\mu\nu}\Biggl\{
-\frac{1}{6}G_{5,X}\left[(\Box\phi)^3-3\Box\phi(\nabla_\alpha\nabla_\beta\phi)^2
+2(\nabla_\alpha\nabla_\beta\phi)^3\right]
+\nabla_\alpha G_5R^{\alpha\beta}\nabla_\beta\phi
\cr&&
-\frac{1}{2}\nabla_\alpha\left(G_{5,\phi}\nabla^\alpha\phi\right)\Box\phi
+\frac{1}{2}\nabla_\alpha\left(G_{5,\phi}\nabla_\beta\phi\right)\nabla^\alpha\nabla^\beta\phi
-\frac{1}{2}\nabla_\alpha G_{5,X}\nabla^\alpha X\Box\phi
\cr&&
+\frac{1}{2}
\nabla_\alpha G_{5,X}\nabla_\beta X\nabla^\alpha\nabla^\beta\phi
-\frac{1}{4}\nabla^\lambda G_{5,X}\nabla_\lambda\phi\left[
(\Box\phi)^2-(\nabla_\alpha\nabla_\beta\phi)^2\right]
\cr&&
+\frac{1}{2}G_{5,X}R_{\alpha\beta}\nabla^\alpha\phi\nabla^\beta\phi \Box\phi
-\frac{1}{2}G_{5,X}R_{\alpha\lambda\beta\rho}
\nabla^\alpha\nabla^\beta\phi\nabla^\lambda\phi\nabla^\rho\phi
\Biggr\},
\end{eqnarray}
\begin{eqnarray}
P_\phi^2&=&K_\phi,
\\
P_\phi^3&=&\nabla_\mu G_{3,\phi}\nabla^\mu\phi,
\\
P_\phi^4&=&G_{4,\phi}R+G_{4,\phi X}\left[(\Box\phi)^2-(\nabla_\mu\nabla_\nu\phi)^2\right],
\\
P_\phi^5&=&-\nabla_\mu G_{5,\phi }G^{\mu\nu} \nabla_\nu\phi -
\frac{1}{6}G_{5,\phi X}
\left[
(\Box\phi)^3-3\Box\phi (\nabla_\mu\nabla_\nu \phi)^2+2(\nabla_\mu\nabla_\nu \phi)^3
\right],
\end{eqnarray}
and
\begin{eqnarray}
J_\mu^2&=&-{\cal L}_{2X}\nabla_\mu\phi,
\\
J_\mu^3&=&-{\cal L}_{3X}
\nabla_\mu\phi+ G_{3,X}\nabla_\mu X+2G_{3,\phi}\nabla_\mu\phi,
\\
J_\mu^4&=&-{\cal L}_{4X}\nabla_\mu\phi+
2G_{4,X}R_{\mu\nu}\nabla^\nu\phi
-2G_{4,XX}\left(\Box\phi\nabla_\mu X-\nabla^\nu X\nabla_\mu\nabla_\nu\phi\right)
\cr&&
-2G_{4,\phi X}\left(\Box\phi\nabla_\mu \phi +\nabla_\mu X\right),
%%%
\\
%%%
J_\mu^5&=&
-{\cal L}_{5X}\nabla_\mu\phi-2G_{5,\phi }G_{\mu\nu}\nabla^\nu\phi
\cr&&
-G_{5,X}\left[G_{\mu\nu}\nabla^\nu X+
R_{\mu\nu}\Box\phi \nabla^\nu\phi-R_{\nu\lambda}\nabla^\nu\phi\nabla^\lambda\nabla_\mu\phi
-R_{\alpha\mu\beta\nu}\nabla^\nu\phi\nabla^\alpha\nabla^\beta\phi
\right]
\cr
&&
+G_{5,XX}\left\{
\frac{1}{2}\nabla_\mu X\left[(\Box\phi)^2-(\nabla_\alpha\nabla_\beta\phi)^2\right]
-\nabla_\nu X\left(\Box\phi \nabla_\mu\nabla^\nu\phi-\nabla_\alpha\nabla_\mu\phi
\nabla^\alpha\nabla^\nu\phi\right)
\right\}
\cr
&&
+G_{5,\phi X}\left\{
\frac{1}{2}\nabla_\mu\phi\left[(\Box\phi)^2-(\nabla_\alpha\nabla_\beta\phi)^2\right]
+\Box\phi\nabla_\mu X-\nabla^\nu X\nabla_\nu\nabla_\mu \phi
\right\}.
\end{eqnarray}

\bibliography{mybiblio}
\bibliographystyle{JHEP}
\end{document}